\documentclass[useAMS,usenatbib]{mn2e}

\usepackage{graphicx}

\title[Electron capture of iron group nuclei in magnetars]{Electron capture of iron group nuclei in magnetars}
\author[Liu Jing-Jing ]{Liu Jing-Jing$^{1}$\thanks{E-mail:liujingjing68@126.com}
\footnotemark[1]\thanks{Project supported by the Advanced Academy Special Foundation of Sanya under Grant No 2011YD14.}\\
$^{1}$College of Science and Technology, Qiongzhou University, Sanya, 572022, China}
\begin{document}

\date{Accepted 2013 March 15. Received 2012 March 14; in original form 2013 January 11}

\pagerange{\pageref{firstpage}--\pageref{lastpage}} \pubyear{2012}

\maketitle

\label{firstpage}

\begin{abstract}
Using the theory of relativity in superstrong magnetic fields (SMFs) and nuclear shell model, we carry out
estimation for electron capture (EC) rates on iron group nuclei in SMFs. The rates of change of electronic
abundance (RCEA) due to EC are also investigated in SMFs. It is concluded that the EC rates of most iron group
nuclides are increased greatly by SMFs and even exceeded by nine orders of magnitude. On the other hand, RCEA is influenced greatly by SMFs and even reduced by more than eight orders of magnitude in the EC reaction.We also compare our results with those of Fuller et al.(FFN) and Aufderheide et al.(AUFD) in the case with and without SMFs. The results show that our results are in good agreement with AUFD's, but the rates of FFN's are about close to one order of magnitude bigger than ours in the case without SMFs. On the contrary, our calculated rates for most nuclides in SMFs are increased and even exceeded as much as for nine and eight orders of magnitude of compared to FFN's and AUFD's results, which is in the case without SMFs, respectively.
\end{abstract}


\section{Introduction}

It is well known that electron captures (EC) on nuclei play an important
role in the dynamics of the collapsing core of a massive star that
leads to a supernova explosion. Near the end of their lives, the
core of a massive star ($M \geq 10M_{\odot}$) consists predominantly
of iron and its nuclear neighbors, the "iron group" elements. The
poignant supernova explosions will lead up due to the unstable
nuclear burning and iron nuclei collapse. During the pre-collapse
phase, EC reduces the number of electrons available for pressure
support. At the same time, the neutrinos produced by EC freely
escape from the star for values of the matter density about $10^{11}
\rm{g/cm^3}$, removing energy and entropy from the core. Some
researches show that the EC, especially for the iron group nucleus
play a crucial role in the late evolution stages of massive stars.
On the other hand, electron-capture supernovae have been proposed as
possible origin of elements beyond iron, in particular of heavy
r-process elements. As a result, the treatment of EC significantly
influences the initial conditions for the entire post-bounce
evolution of the supernova.

Under the supernova like explosion conditions, the problem of EC has
been an active area of investigations for many years. \citet{b8,b9};
\citet{b6} investigated the EC rates of many iron group nucleuses.
In the same environment, \citet{b1, b2}; \citet{b13}also discussed
the EC rates of the iron group nucleus, but their discussions are
for the case without a strong magnetic fields(hereafter SMFs).
\citet{b7, b14} discussed the influence of
 SMFs on EC at zero temperature in neutron stars, which they focused only on the ground state transition of the simple nuclei and paid no attention to the Gamow--Teller (GT) transition and a number of iron group nuclei.Due to the importance of SMFs in astrophysical environments,
\citet{b15, b16}; \citet{b17, b18}; \citet{b20, b21, b22, b23, b24, b25} did some works on EC and especially discussed the weak interaction reactions and neutrino energy loss on iron group nucleus.

Some studies show that on the surface of neutron stars, the strengths of magnetic fields are in the range of $10^{8}G$ to
$10^{13}G$. Especially for some magnetars, the range of SMFs strengths is from
$10^{13}G$ to $10^{15}G$ \citep{b19, b21, b24, b25, b27, b28}. For such SMFs, the classical description of the trajectories of a free electron is no longer valid and quantum effect must be considered. How would the SMFs effect on the EC in magnetars? How would the SMFs effect on the cooling system in magnetars? How would the EC affect on the evolution in magnetars? How would the SMFs affect on the Fermi energy and the electron chemical potential? How would the SMFs affect on the GT transition in the EC process? These are very interesting and challenging problems in magnetars.

Previous works \citep{b15, b17, b18, b19} show that SMFs effect on
electron capture rate and neutrino energy loss rates greatly and
decreases with the increasing of the strength of magnetic field.
Recent studies\citep{b21, b24, b25, b27, b28} have found,
strengthened the magnetic field will make the Fermi surface would
elongate from a spherical surface to a Landau surface along the
magnetic field direction and  its level is perpendicular to the
magnetic field direction and quantized.  Thus, we have to adapt the
theory of non-relativistic Landau level.we will discuss the level
density in detail in SMFs beause the calculation on EC is strongly
depended on the choice of level density.

Basing on the nuclear $pf$--shell model and theory of relativity in
SMFs \citep{b27, b28,b11} in this paper, we focus on the iron group
nuclei and investigate their EC rates. We also discuss the rates of
change of electronic abundance (RCEA) due to EC in SMFs.

\section[]{The study of EC in an SMFs}

An SMFs is considered along the z-axis according to theory of
relativity in superstrong magnetic fields. The Dirac equation can be
solved exactly. The positive energy levels of an electron in an SMFs
are given by \citep{b27, b28, b21, b24, b25}
\begin{equation}
   \frac{E_n}{m_e c^2}=[(\frac{p_z}{m_e c})+1+2(n+\frac{1}{2}+\sigma)b]^{1/2},(n=0,1,2,3....)
\end{equation}
 where
 $b=\frac{B}{B_{cr}}=0.02266B_{12}$; $B=10^{12}B_{12}$; $B_{cr}=\frac{m^2_e c^3}{e\hbar}=4.414\times10^3G$ and $p_z$is
 the electron momentum along the field, $\sigma$ is the spin quantum number of an electron, $\sigma=1/2$ when $n=0$; $\sigma=\pm1/2$ when $n\geq1$.

 In an extremely strong magnetic field $(B\gg B_{cr})$, the Landau column becomes a very long and very narrow cylinder
 along the magnetic field, the electron chemical potential is found by inverting the expression for the lepton number density\citep{b21, b23}
 \begin{equation}
n_e=\rho
Y_e=\frac{b}{2\pi^2\lambda_e^3}\sum_{0}^{\infty}q_{n0}\int^{\infty}_{0}(f_{-e}-f_{+e})dp_z,
\end{equation}
where $\rho$ is the mass density in $g/cm^3$; $Y_e=\frac{Z}{A}$ is the electron fraction; $\lambda_e=\frac{h}{m_e c}$ is the Compton wavelength, $m_e$ is the electron mass and  $c$ is the light speed. $q_{n0}=2-\delta_{n0}$is the electron degenerate number, $f_{-e}=[1+exp(\frac{\varepsilon_n-U_F-1}{kT})]^{-1}$ and $f_{+e}=[1+exp(\frac{\varepsilon_n+U_F+1}{kT})]^{-1}$are the electron and positron distribution functions respectively, $k$ is the Boltzmann constant, $T$ is the electron temperature and $U_F$
is the electron chemical potential.

We naturally draw a conclusion that the stronger the magnetic field,
the higher the Fermi energy of electrons. Unfortunately, some works
\citep{b27, b28} show that currently the most popular viewpoint
which is the stronger the magnetic field, the lower the electron
Fermi energy, is completely contrary to above conclusion. In an
extremely strong magnetic field, the electrons are in disparate
energy states in order one by one from the lowest energy state up to
the Fermi energy with the highest momentum according to the Pauli
exclusion principle. The electron energy state in a unit volume
should be equal to the electron number density. Thus, the electron
Fermi energy is also determined by \citep{b11}
\begin{eqnarray}
\rho Y_e&=&\frac{3\pi}{bN_A }(\frac{m_e c}{h})^3(\gamma_e)^4\int^{1}_{0}(1-\frac{1}{(\gamma_e)^2}-\chi^2)^{\frac{3}{2}}d\chi\nonumber\\
&&-\frac{2\pi\gamma_e}{N_A }(\frac{m_ec}{h})^3\sqrt{2b},
\end{eqnarray}
where $N_A$ is Avogadro constant; $\gamma_e$ and $\chi$ are two non-dimensional variables and defined as $\gamma_e=\frac{E_F(e)}{m_ec^2}$, $\chi=\frac{P_Zc}{E_F(e)}$, respectively; $E_F(e)$is the electron Fermi energy.

In the case without SMFs, the electron capture rates for the $k$ th nucleus $(Z,A )$ in thermal equilibrium at temperature $T$ is given by a sum over the initial parent states $i$ and the final daughter states $f$ \citep{b8, b9, b17, b18}
\begin{equation}
 \lambda_{k}=\lambda_{ec}=\sum_{i}\frac{(2J_i+1)e^{\frac{-E_i}{kT}}}{G(Z,A,T)}
 \sum_{f}\lambda_{if},
\end{equation}
where  $J_i$ and $E_i$ are the spin and excitation energies of the
parent states, $G(Z,A,T)$  is the nuclear partition function and given by
\begin{equation}
\label{eq.2} G(Z,A,T)=\sum_i(2J_i+1)exp(-\frac{E_i}{kT}).
\end{equation}

According to the level density formula, when the contribution from the excite states is discussed, the nuclear partition function approximately becomes\citep{b1}
\begin{eqnarray}
\label{eq.3}
G(Z,A,T) &\approx& (2J_0+1)+ \int_0^\infty dE \int _{J,\pi}dJd\pi (2J_i+1)\nonumber\\
&&\times \vartheta(E, J, \pi)exp(-\frac{E_i}{kT}),
\end{eqnarray}
where the level density  is given by \citep{b31}
\begin{equation}
\label{eq.4} \vartheta(E, J,
\pi)=\frac{1}{\sqrt{2\pi}\sigma}\frac{\sqrt{\pi}}{12a^{\frac{1}{4}}}\times\frac{exp[2\sqrt{a(E-\delta)}]}{(E-\delta)^{\frac{5}{4}}}
f(E, J, \pi),
\end{equation}
where
\begin{equation}
\label{eq.5} f(E, J,
\pi)=\frac{1}{2}\frac{(2J+1)}{2\sigma^2}exp[-\frac{J(J+1)}{2\sigma^2}],
\end{equation}
where $a$ is the level density parameter, $\delta$ is the backshift (pairing correction).  $\sigma$ is defined as
\begin{equation}
\label{eq.6}
\sigma=(\frac{2m_uAR^2}{2\hbar^2})^{\frac{1}{2}}[\frac{(E-\delta)}{a}]^{\frac{1}{4}},
\end{equation}
where $R$ is the radius and $m_u=\frac{1}{N_A}$ is the atomic mass unit.

The EC rate from one of the initial states to all possible final
states is $\lambda_{if}$,
$\lambda_{if}=\frac{\ln2}{(ft)_{if}}f_{if}$ with the relation
$\frac{1}{(ft)_{if}}=\frac{1}{(ft)_{if}^{F}}+\frac{1}{(ft)_{if}^{GT}}$
.The $ft$ -values and the corresponding GT or Fermi transition
matrix elements are related by the following expression
\begin{equation}
 \frac{1}{(ft)_{if}}=\frac{1}{(ft)_{if}^{F}}+\frac{1}{(ft)_{if}^{GT}}=\frac{10^{3.79}}{|M_F|^2_{if}}+\frac{10^{3.596}}{|M_{GT}|^2_{if}},
\end{equation}

The Fermi matrix element and the GT matrix element are given as follows respectively \citep{b8, b9}
\begin{eqnarray}
 |M_F|^2&=&\frac{1}{2J_i+1}\sum_{m_i}\sum_{m-f}|\langle \psi_f m_f|\sum_N \tau_N^-|\psi_im_i\rangle|^2\nonumber\\
 &=&T'(T'+1)-T_Z^{'i}(T_Z^{'i}-1),
\end{eqnarray}

\begin{equation}
 |M_{GT}|^2=\frac{1}{2J_i+1}\sum_{m_i}\sum_{m-f}|\langle \psi_f m_f|\sum_N
 \tau_N^-\kappa_N|\psi_im_i\rangle|^2,
\end{equation}
where $T'$ is the nuclear isospin and  $T'_Z=T^{'i}_Z=(Z-N)/2$ is its projection for the parent or the daughter nucleus. $|\psi_i m_i\rangle$ is the initial parent state,  $\langle \psi_f m_f|$ is the finial daughter state, and the Fermi matrix element is averaged over initial and summed over finial nuclear spins. $\sum_N \tau_N^-$ is he minus component of isovector, spatial scalar operator  $T^{'-}$ which commuter with the total isospin $T^{'2}$ . $\kappa$ is the Pauli spin operator and $\sum_N \tau_N^- \kappa_N$ is a spatial vector and an isovector.

The total amount of Gamow-teller(GT) strength
available for an initial state is $S_{GT^+}$ and given by summing over a complete
set o final states in GT transition matrix elements
$|M_{GT}|^{2}_{if}$. The Shell Model Monte Carlo(SMMC) method is also used to
calculate the response function $R_A(\tau)$ of an operator $\hat{A}$
at an imaginary-time $\tau$. By using a spectral distribution of
initial and final states $|i\rangle$ and $|f\rangle$ with energies
$E_i$ and $E_f$. $R_A(\tau)$ is given by \citep{b13}
\begin{equation}
R_A(\tau)=\frac{\sum_{if}(2J_i+1)e^{-\beta E_i}e^{-\tau
(E_f-E_i)}|\langle f|\hat{A}|i\rangle|^2}{\sum_i (2J_i+1)e^{-\beta
E_i}},
\label{eq:008}
\end{equation}
Note that the total strength for the operator is given by
$R(\tau=0)$. The strength distribution is given by
\begin{eqnarray}
S_{GT^+}(E) &=& \frac{\sum_{if}\delta (E-E_f+E_i)(2J_i+1)e^{-\beta
E_i}|\langle f|\hat{A}|i\rangle|^2}{\sum_i (2J_i+1)e^{-\beta E_i}} \nonumber\\
&=& S_{A}(E),
 \label{eq:009}
\end{eqnarray}
which is related to $R_A(\tau)$ by a Laplace Transform,
$R_A(\tau)=\int_{-\infty}^{\infty}S_A(E)e^{-\tau E}dE$. Note that
here $E$ is the energy transfer within the parent nucleus, and that
the strength distribution $S_{GT^+}(E)$ has units of $\rm
{MeV^{-1}}$ and $\beta=\frac{1}{T_N}$, $T_N$ is the nuclear
temperature.

We can find the phase space factor $f_{if}$ in SMF from Refs \citep{b16,b17}, and it is defined as
\begin{equation}
f_{if}^B=\frac{b}{2}\sum_{0}^{\infty}\theta_n
=\frac{b}{2}\sum_{0}^{\infty}
q_{no}\int_{q_n}^{\infty}(Q_{if}+\varepsilon_n)^2
F(Z,\varepsilon_n)f_{-e}dp,
\end{equation}
where $Q_{if}=Q_{00}+E_i-E_f$, is the EC threshold energy;
$Q_{00}=M_p c^2-M_d c^2$ , with $M_p$ and $M_d$ being the masses of
the parent nucleus and the daughter nucleus respectively; $E_i$ and
$E_f$ are the excitation energies of the  $i$ th states and  $f$ th
state of the nucleus respectively; the  $\varepsilon_n$ is the total
rest mass and kinetic energies; $F(Z, \varepsilon_n)$ is the Coulomb
wave correction which is the ratio of the square of the electron
wave function distorted by the coulomb scattering potential to the
square of wave function of the free electron. We assume that a SMFs
will have no effect on $F(Z, \varepsilon_n)$, which is valid only
under the condition that the electron wave-functions are locally
approximated by the plane-wave functions. \citep{b7} The condition
requires that the Fermi wavelength $\lambda_F\sim\frac{\hbar}{P_F}$
($P_F$ is the Fermi momentum without a magnetic field) be smaller
than the radius
 $\sqrt{2} \zeta$ (where $\zeta=\frac{\lambda_e}{b}$ ) of the cylinder which corresponds to the lowest Landau level. \citep{b4}

The $q_n$ is defined as
\begin{equation}
q_n=\left\lbrace \begin{array}{ll}~\sqrt{Q^2_{if}-\Theta},~~~~~~( Q_{if}<\Theta^{1/2})\\
                                  ~0 ~~~~~~~~~~~~~~~~~~~(otherwise).
                             \end{array} \right.
\end{equation}
where $\Theta=1+2(n+\frac{1}{2}+\sigma)b$.

All of these factors are considered, the electron capture rates for the $k$ th nucleus $(Z,A )$ in thermal equilibrium at
temperature $T$ in SMFs is given by
\begin{eqnarray}
 \lambda_{ec}^B &=& \sum_{i}\frac{(2J_i+1)e^{\frac{-E_i}{kT}}}{G(Z,A,T)} \sum_{f}\lambda_{if}^B \nonumber\\
 &=&\sum_{i}\frac{(2J_i+1)e^{\frac{-E_i}{kT}}}{G(Z,A,T)}
 \sum_{f}\frac{\ln2}{(ft)_{if}}f_{if}^B.
\end{eqnarray}

The Shell Model Monte Carlo(SMMC) method is used to discuss the total amount of Gamow-teller(GT) strength in the process of
our EC calculation. Thus, there are some difference among our rates, FFN's and AFUD's. In order to compared our results with
those of FFN's and AUFD's in the case without SMFs, the scale factor $k_1$ and $k_2$ are defined as follows
\begin{equation}
k_1=\frac{\lambda_{ec}(\rm{FFN})}{\lambda_{ec}^{B_{12}=0}(\rm{LJ})},
\label{120}
\end{equation}
\begin{equation}
k_2=\frac{\lambda_{ec}(\rm{AUFD})}{\lambda_{ec}^{B_{12}=0}(\rm{LJ})}.
\label{121}
\end{equation}

On the other hand, what really matters for stellar evolution is the electronic abundance
$Y_e$, the rate of change of electronic abundance (RCEA) caused by
each nucleus. Therefore, the RCEA due to EC on the $k$ th nucleus is
very important in SMFs. It is given by
\begin{equation}
\dot{Y^{ec}_{e}}(k)=\frac{dY_e}{dt}=-\frac{X_k}{A_k}\lambda^{ec}_k,
\label{120}
\end{equation}
where $X_k$ is the mass fraction of the $k$ th nucleus and $A_k$ is the mass number of the $k$ th nucleus.

\section{The EC rates of iron group nucleus in SMFs and disscusion}

The electron chemical potential is a very sensitivity parameter in
the process of EC. Figures 1--5 show the EC rates of some iron group
nuclei as a function of $U_F$  under the condition of $\rho
Y_e=5.86\times10^8 g/cm^3$; $\rho Y_e=3.3\times10^{10} g/cm^3$ and
$T_9=5$ ( $T_9$ is the temperature in units of $10^9K$). One sees
that at relative lower density the electron chemical potential has a relative minor effect on
the EC rates for most nuclides. But the EC rates of most
nuclides are influenced greatly at relative higher density.
For example, for most iron group nuclei (e.g.$^{55-60}$Co and
$^{56-63}$Ni), the EC rates increased no more than by
four and six orders of magnitude at $T_9=5$, $\rho Y_e=5.86\times10^8 g/cm^3$ and $\rho Y_e=3.3\times10^{10} g/cm^3$, respectively.

On the other hand, one can see that as increasing of SMFs, the
higher the density, the smaller the influence on EC is, because the
electron energy and $U_F$ are very large at higher density
surroundings and the higher Landau levels are occupied by electrons.
According to theory of relativity in SMFs, with increasing of the
magnetic fields at relative higher density, $U_F$ increases greatly,
and the lower SMFs effects. It is because that in an extremely
strong magnetic field( $(B\gg B_{cr})$ ), the Landau column becomes
a very long and very narrow cylinder along the magnetic field, the
electron chemical potential and electron energy are so strongly
dependent on the density. For example, according our calculations,
with increasing of the magnetic fields from $10^{13}G$ to
$10^{18}G$, the electron chemical potential increases from $4.5 m_e
c^2$ to
 $13.5 m_e c^2$ at $\rho_7=220, Y_e=0.44$  and from
$110 m_e c^2$ to $390 m_e c^2$ at $\rho_7=5.43\times10^8, Y_e=0.41$ ( $\rho_7$ is the density in units of $10^7 g/cm^3$ )

\begin{figure}
\centering
    \includegraphics[width=4cm,height=4cm]{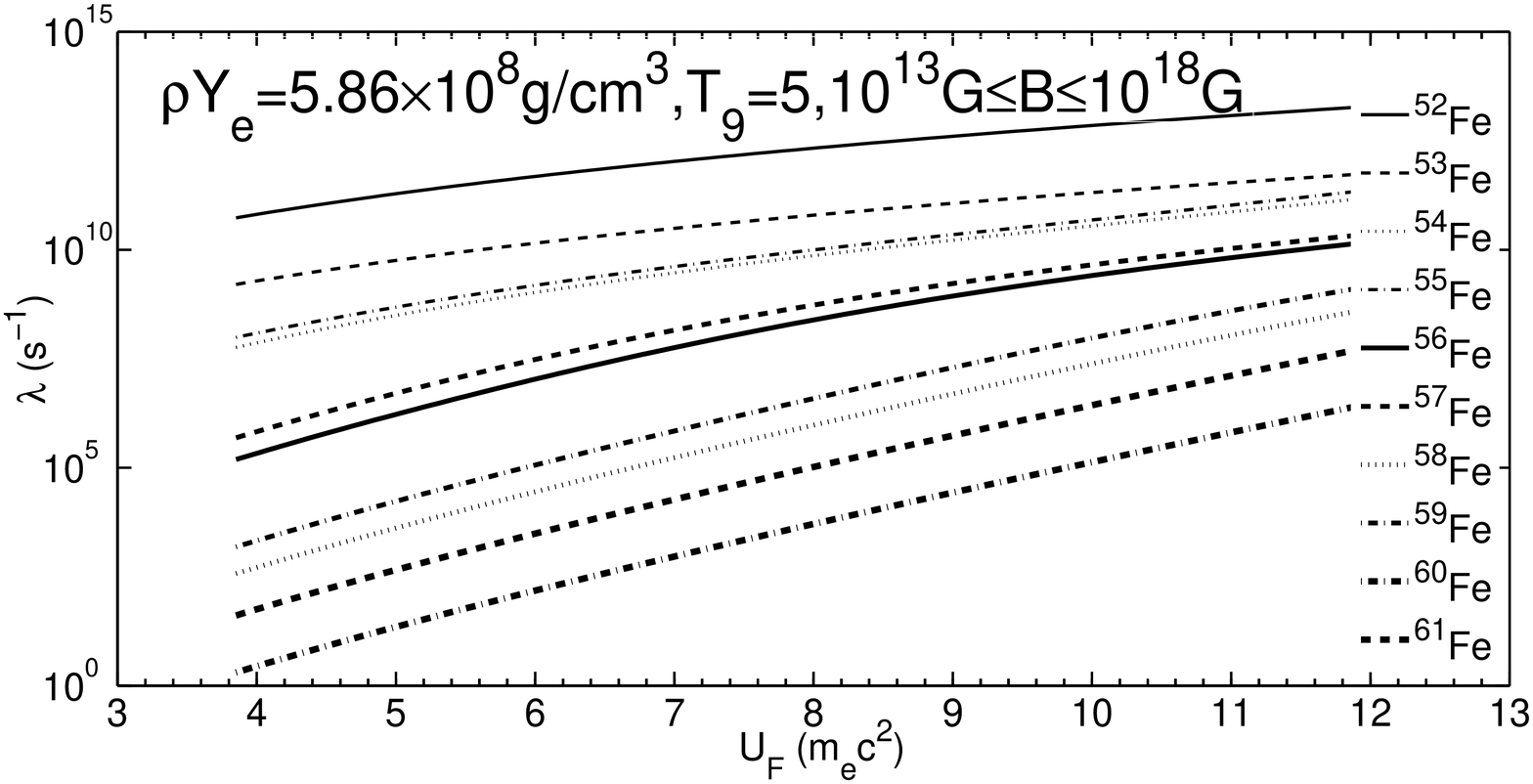}
    \includegraphics[width=4cm,height=4cm]{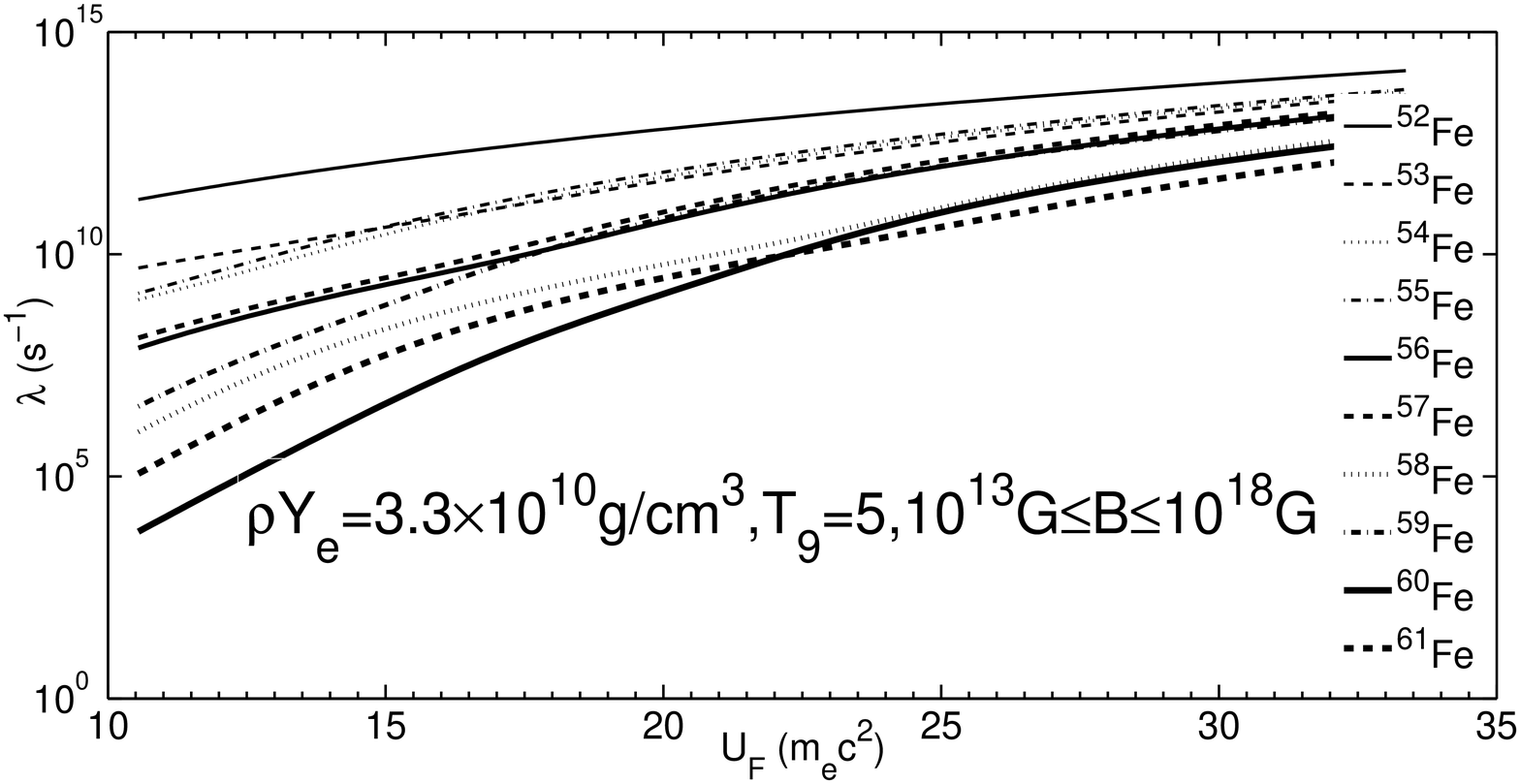}
   \caption{The EC rates for $^{52-61}$ Fe as a function of the electron chemical potential at the density  and temperature of
 $\rho Y_e=5.86\times10^8 g/cm^3$; $\rho Y_e=3.3\times10^{10} g/cm^3$ and $T_9=5$}
   \label{Fig:1}
\end{figure}

%
\begin{figure}
\centering
    \includegraphics[width=4cm,height=4cm]{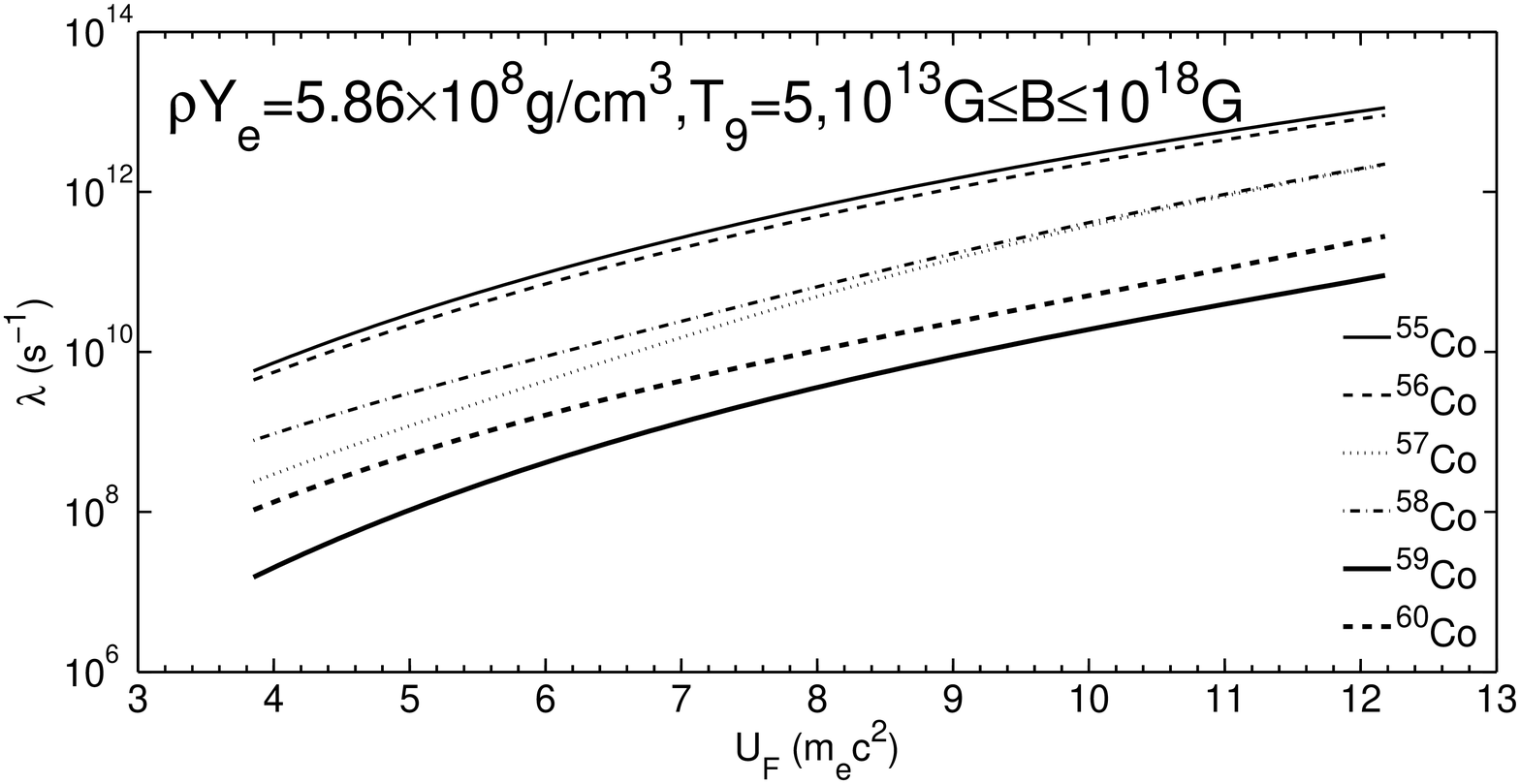}
    \includegraphics[width=4cm,height=4cm]{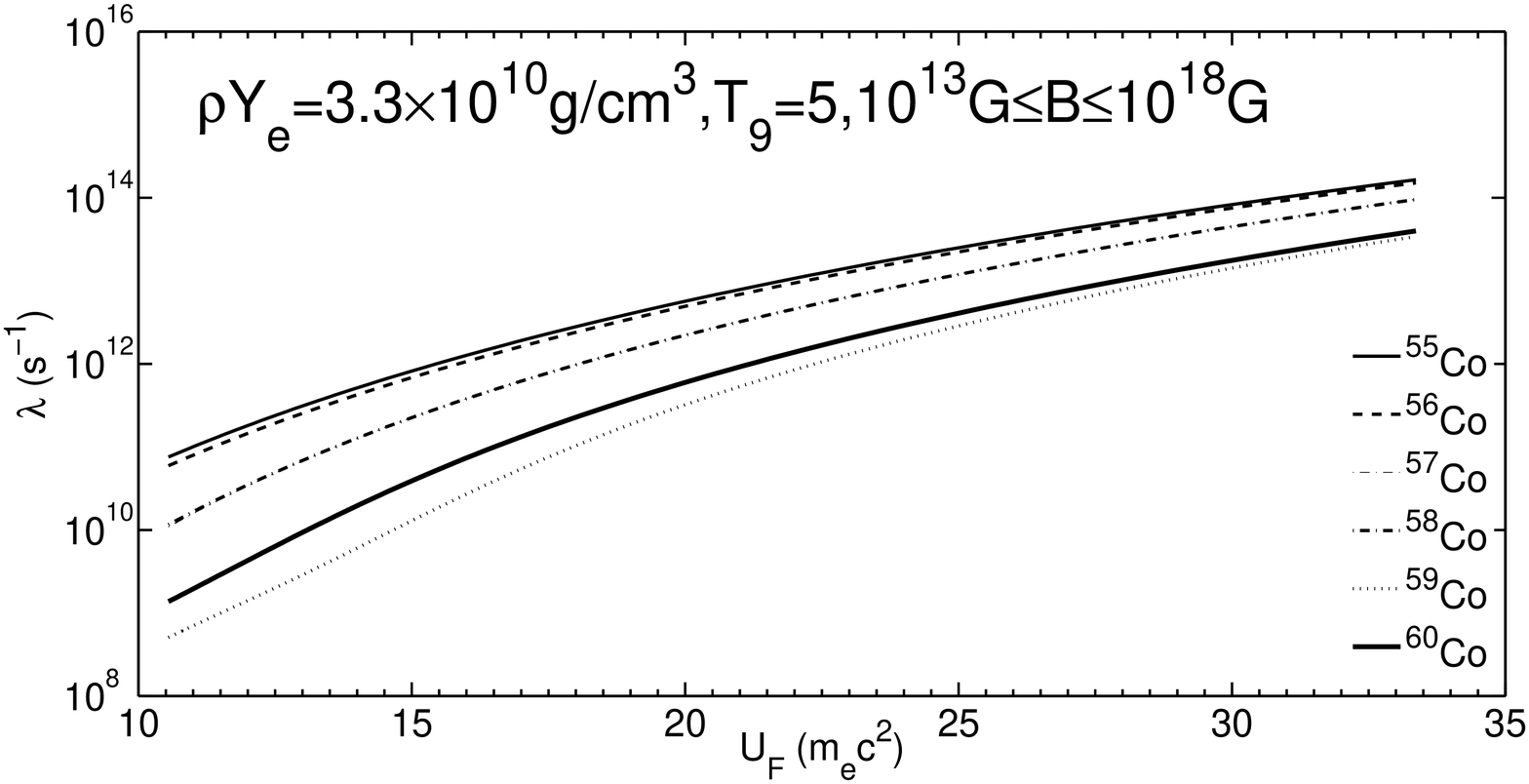}
   \caption{The EC rates for $^{55-60}$ Co as a function of the electron chemical potential at the density
   and temperature of $\rho Y_e=5.86\times10^8 g/cm^3$; $\rho Y_e=3.3\times10^{10} g/cm^3$ and $T_9=5$}
   \label{Fig:2}
\end{figure}
%
\begin{figure}
\centering
    \includegraphics[width=4cm,height=4cm]{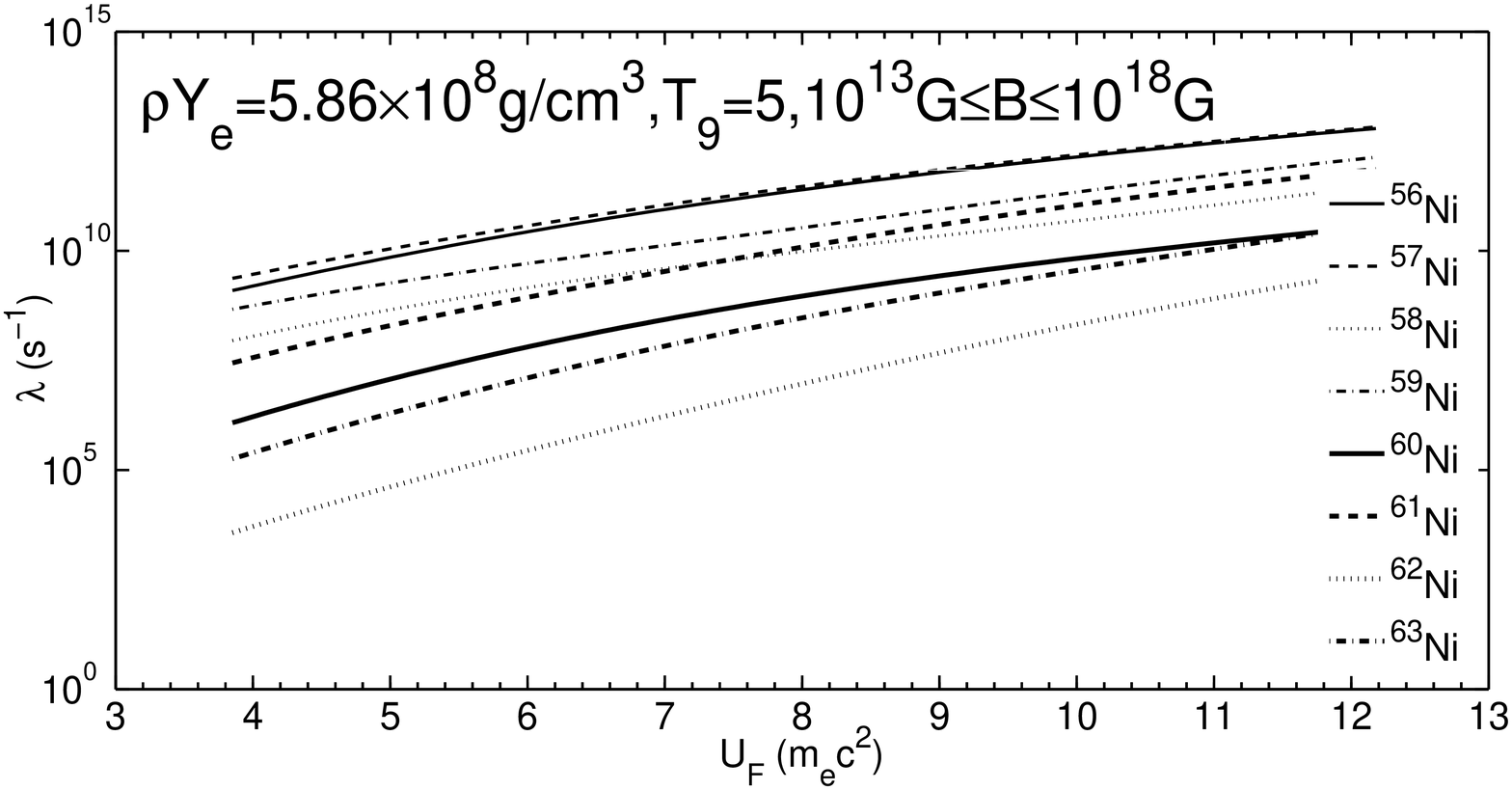}
    \includegraphics[width=4cm,height=4cm]{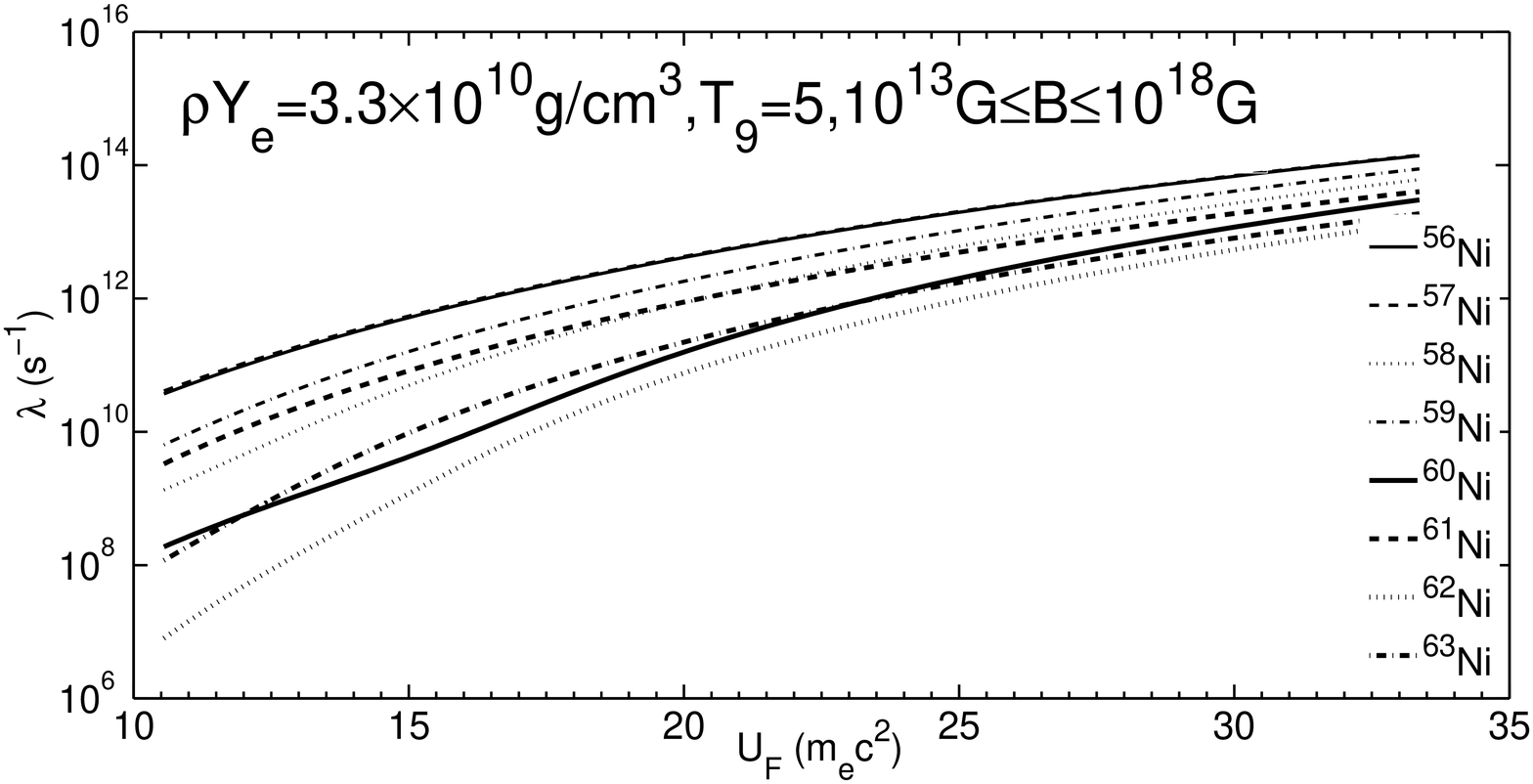}
   \caption{The EC rates for $^{56-63}$ Ni as a function of the electron chemical potential at the density
   and temperature of $\rho Y_e=5.86\times10^8 g/cm^3$; $\rho Y_e=3.3\times10^{10} g/cm^3$ and $T_9=5$}
\end{figure}

\begin{figure}
\centering
    \includegraphics[width=4cm,height=4cm]{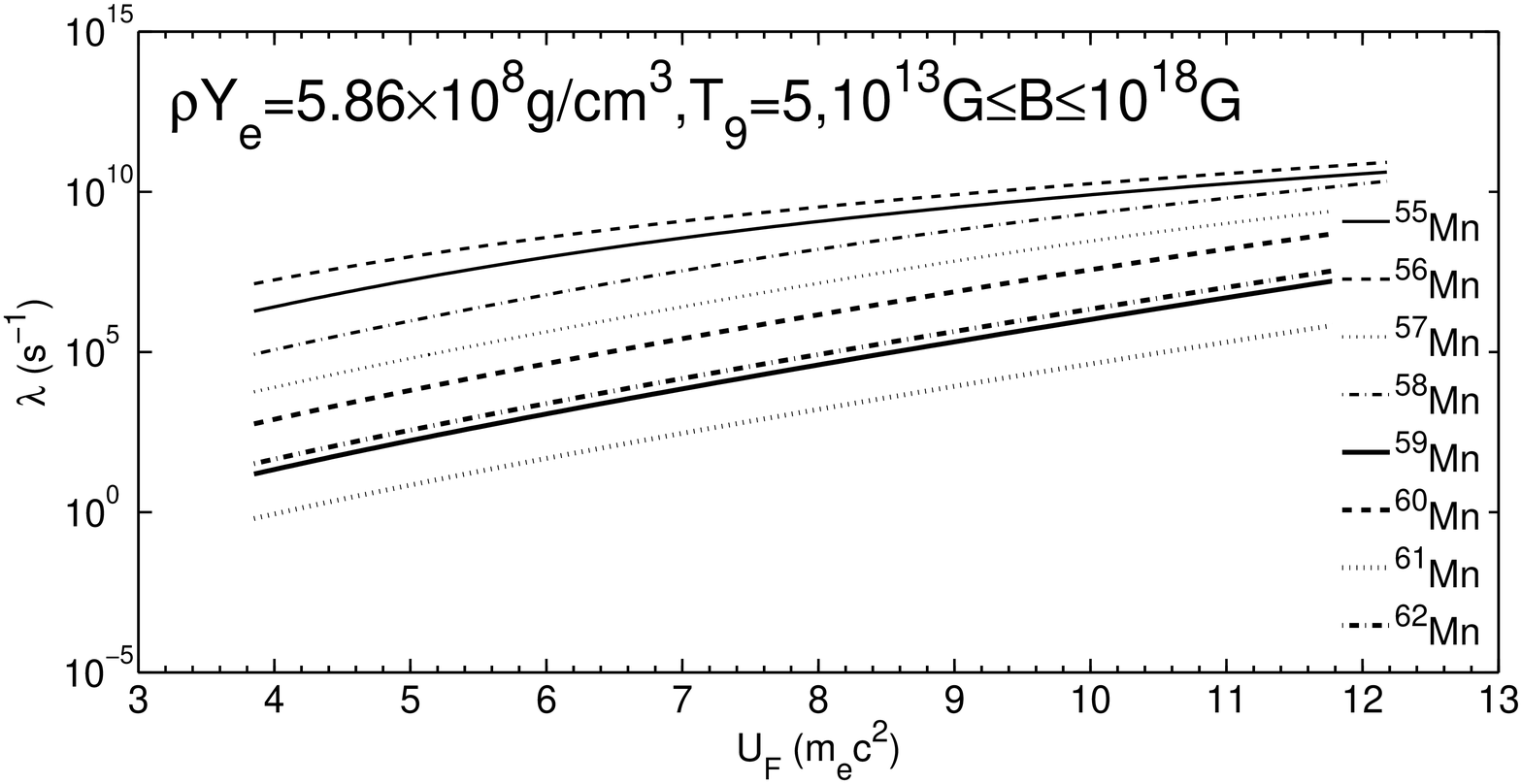}
    \includegraphics[width=4cm,height=4cm]{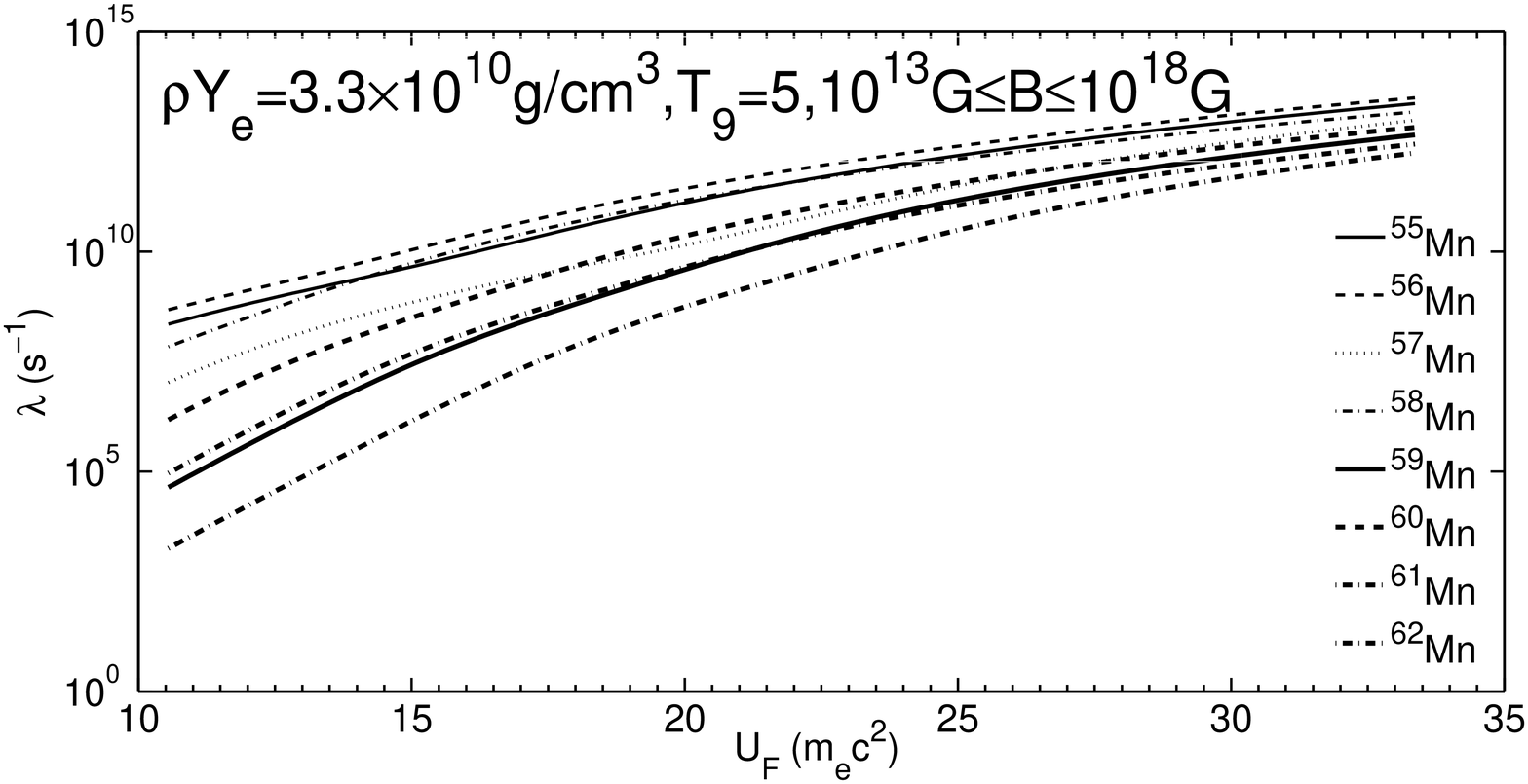}
   \caption{The EC rates for $^{55-62}$ Mn as a function of the electron chemical potential at the density
   and temperature of $\rho Y_e=5.86\times10^8 g/cm^3$; $\rho Y_e=3.3\times10^{10} g/cm^3$ and $T_9=5$}
   \label{Fig:4}
\end{figure}

\begin{figure}
\centering
    \includegraphics[width=4cm,height=4cm]{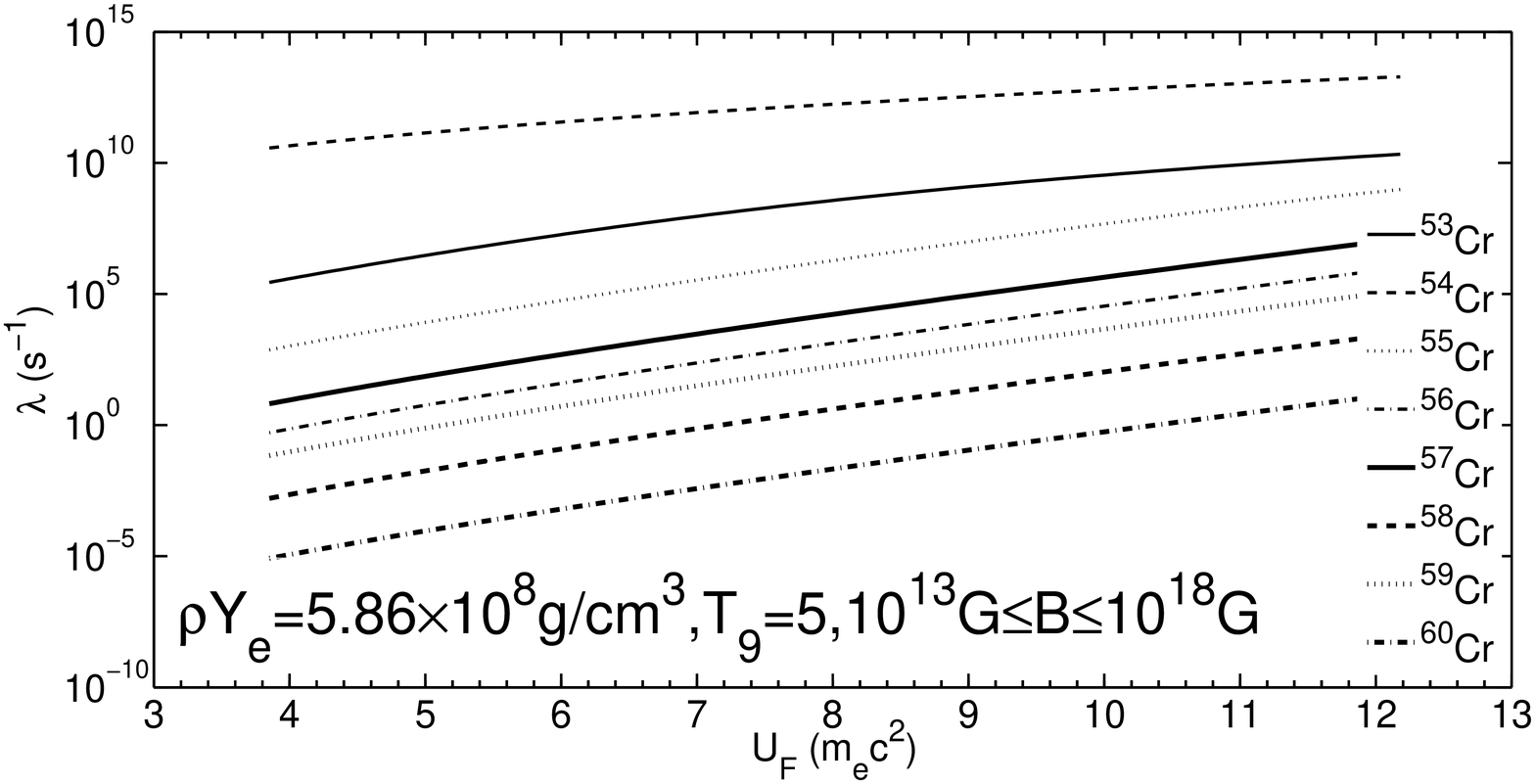}
    \includegraphics[width=4cm,height=4cm]{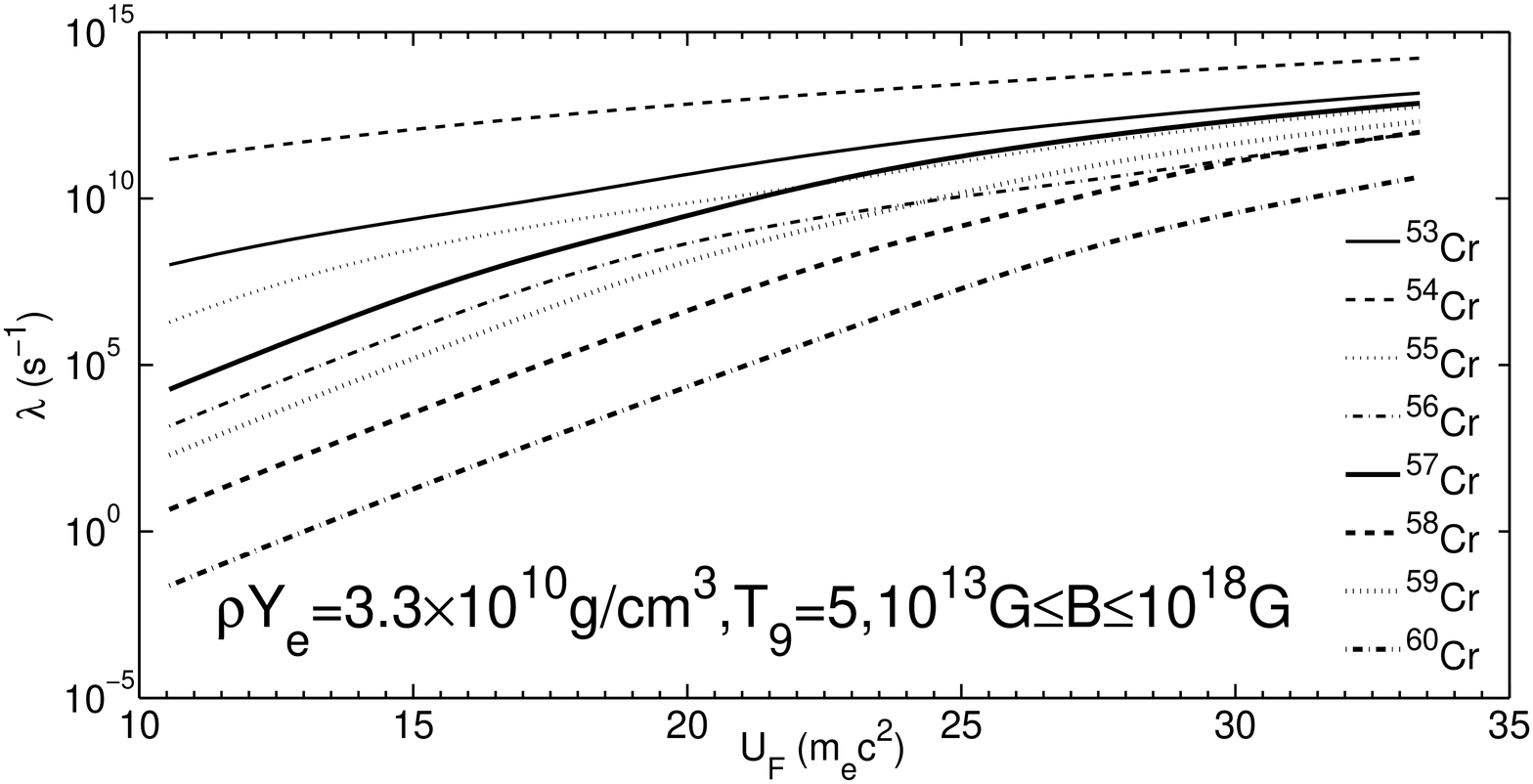}
   \caption{The EC rates for $^{53-60}$ Cr as a function of the electron chemical potential at the density
   and temperature of $\rho Y_e=5.86\times10^8 g/cm^3$; $\rho Y_e=3.3\times10^{10} g/cm^3$ and $T_9=5$}
   \label{Fig:5}
\end{figure}

Figures 6--10 show the EC rates of some iron group nuclei as a
function of the magnetic fields $B$  under the condition of $\rho
Y_e=3.30\times10^{10} g/cm^3$; $\rho Y_e=1.13\times10^{14} g/cm^3$
and $T_9=9,15$ respectively. We find that the EC rates of most
nuclides influenced greatly at relative lower temperature and
density. For example, for most iron group nuclei (e.g. $^{55-60}$ Co
and $^{56-63}$Ni ), the EC rates increase and even exceed by four
orders of magnitude at $\rho Y_e=3.30\times10^{10} g/cm^3$ and
$T_9=9$. However, at $\rho Y_e=1.13\times10^{14} g/cm^3$ and
$T_9=15$ the increase is no more than by three orders of magnitude.

Comparing the results in these figures, it can be seen that the GT
transition process of the EC process may not be dominant at lower
temperature. This process is dominated by the low-energy transition.
Therefore, the effect produced by this kind of densities is not very
obvious in SMFs. We find that the distribution of the electron gas
with high temperature and high density must satisfy the Fermi-Dirac
distribution. The GT transition strength of nuclide is distributed
in the form of the centrosymmetric Gaussian function about the GT
resonance point. So the energies of the electrons taking part in the
GT resonance transitions in the high--energy range are not symmetric
in SMFs. The distribution of Gaussian function increases and includes
more electrons to take part in the electron capture reactions.
Therefore, with increasing of the temperature, it obviously
accelerates the progress of the electron capture process. It
inevitably leads to great increases of the EC rates.


\begin{table*}
\caption{Comparisons of our calculations $\lambda^B_{ec}(\rm{LJ})$ in and not in SMFs for some typical iron group nuclides
with those of FFN's ($\lambda^0_{ec}(\rm{FFN}))$ and AUFD's ($\lambda^0_{ec}(\rm{AUFD}))$, which are for the case without SMFs at
 $\rho_7=5.86, Y_e=0.47, T_9=3.4$.}
\begin{center}
\tiny
\begin{minipage}{140mm}
\begin{tabular}{lllllllll}
\hline
\multicolumn{4}{r}{$\lambda^B_{ec}(\rm{LJ})$} \\
\cline{4-7}
Nuclide &$\lambda^0_{ec}$(FFN)  &$\lambda^0_{ec}$(AUFD) &$B_{12}=0$  &$B_{12}=10$ &$B_{12}=10^3$ &$B_{12}=10^6$ &$k1$ &$k2$\\
\hline
$^{56}$Ni   & 1.30e-2  &1.60e-2  & 1.250e-2 &3.9309e1   &3.0078e3   &1.2274e7  &1.0400 &1.2800\\
$^{57}$Ni   & 9.93e-3  &1.94e-2  & 1.573e-2 &1.1020e1   &4.5929e3   &1.2513e7  &0.6313 &1.2333\\
$^{58}$Ni   & 3.72e-4  &6.36e-4  & 5.878e-4 &1.6875e-1  &1.9710e1   &5.1269e5  &0.6329 &1.0820\\
$^{59}$Ni   & 4.31e-3  &4.37e-3  & 4.146e-3 &2.2885e0   &7.3434e3   &7.5878e6  &1.0396 &1.0540\\
$^{60}$Ni   & 9.17e-6  &1.49e-6  & 1.287e-6 &7.6940e-3  &4.2441e0   &2.4165e3  &7.1251 &1.1577\\
$^{53}$Fe   & 3.91e-2  &2.04e-2  & 1.889e-2 &9.9095e1   &2.2364e3   &3.1121e6  &2.0699 &1.0799\\
$^{54}$Fe   & 2.95e-4  &3.11e-4  & 2.868e-4 &7.4556e0   &1.3858e2   &3.7685e4  &1.0286 &1.0844\\
$^{55}$Fe   & 1.57e-3  &1.61e-3  & 1.357e-3 &2.1021e0   &2.0968e3   &4.2672e6  &1.1570 &1.1864 \\
$^{55}$Co   & 1.36e-1  &1.41e-1  & 1.336e-1 &1.3468e3   &1.2935e4   &5.3337e6  &1.0180 &1.0554\\
$^{56}$Co   & 6.91e-2  &7.40e-2  & 7.026e-2 &1.6156e1   &9.2840e3   &1.3409e7  &0.9835 &1.0532\\
$^{57}$Co   & 3.50e-3  &1.89e-3  & 7.026e-2 &1.6156e1   &9.2840e3   &1.3409e7  &2.1059 &1.1372\\
$^{58}$Co   & 9.93e-3  &1.94e-2  & 1.680e-2 &4.2633e2   &1.1873e4   &8.2574e6  &0.5911 &1.1548\\
\hline
\end{tabular}
\end{minipage}
\end{center}
\end{table*}
\begin{table*}
\caption{Comparisons of our calculations $\lambda^B_{ec}(\rm{LJ})$ in and not in SMFs for some typical iron group nuclides
with those of FFN's ($\lambda^0_{ec}(\rm{FFN}))$ and AUFD's ($\lambda^0_{ec}(\rm{AUFD}))$, which are for the case without SMFs at
$\rho_7=14.5, Y_e=0.45, T_9=3.80$.}
\begin{center}
\tiny
\begin{minipage}{140mm}
\begin{tabular}{lllllllll}
\hline
\multicolumn{4}{r}{$\lambda^B_{ec}(\rm{LJ})$} \\
\cline{4-7}
Nuclide &$\lambda^0_{ec}$(FFN)  &$\lambda^0_{ec}$(AUFD) &$B_{12}=0$ &$B_{12}=10$ &$B_{12}=10^3$ &$B_{12}=10^6$ &$k1$ &$k2$\\
\hline
$^{57}$Co   & 1.04e-2   &1.29e-2    &1.0123e-2  &1.0054e1   &1.1138e3   &1.5740e6  &1.0274 &1.2743\\
$^{58}$Co   & 1.55e-2   &3.07e-2    &2.5604e-2  &4.8841e1   &1.9623e3   &1.5610e6  &0.6054 &1.1990\\
$^{59}$Co   & 5.44e-4   &6.57e-4    &4.3564e-4  &1.5203e-1  &1.1256e1   &6.5784e5  &1.2487 &1.5081\\
$^{60}$Co   & 1.15e-2   &1.27e-2    &1.0012e-2  &3.5286e1   &3.8400e3   &7.1148e5  &1.1486 &1.2685\\
$^{55}$Mn   & 2.03e-5   &2.25e-5    &2.1135e-5  &6.4984e-3  &2.7884e1   &4.8910e4  &0.9605 &1.0646\\
$^{56}$Mn   & 4.29e-5   &2.56e-4    &2.0552e-4  &1.2549e-1  &1.0288e1   &6.1896e4  &0.2087 &1.2456\\
$^{55}$Fe   & 1.21e-2   &6.200e-3   &5.7883e-3  &3.1168e0   &3.6345e2   &9.2804e5  &2.0904 &1.0711\\
$^{56}$Fe   & 2.81e-5   &1.31e-6    &1.2452e-6  &2.2436e-3  &3.0910e-1  &4.1959e4  &22.567 &1.0520\\
$^{57}$Fe   & 4.83e-5   &1.84e-5    &1.6534e-5  &5.8381e-2  &6.4230e0   &4.7899e5  &2.9213 &1.1129\\
$^{60}$Ni   & 1.39e-4   &2.74e-5    &2.3372e-5  &3.4484e-2  &2.0088e0   &6.2462e5  &5.9473 &1.1723\\
$^{61}$Ni   & ......    &1.20e-3    &1.0028e-3  &3.0760e0   &2.3054e2   &6.6119e5  &...... &1.1966 \\
$^{53}$Cr   & 1.63e-5   &2.46e-6    &2.3225e-6  &4.8821e-3  &5.6122e-1  &3.3553e4  &7.0183 &1.0592\\
\hline
\end{tabular}
\end{minipage}
\end{center}
\end{table*}
\begin{table*}
\caption{Comparisons of our calculations $\lambda^B_{ec}(\rm{LJ})$ in and not in SMFs for some typical iron group nuclides
with those of FFN's ($\lambda^0_{ec}(\rm{FFN}))$ and AUFD's ($\lambda^0_{ec}(\rm{AUFD}))$, which are for the case without SMFs at
$\rho_7=106, Y_e=0.43, T_9=4.93$.}
\begin{center}
\tiny
\begin{minipage}{140mm}
\begin{tabular}{lllllllll}
\hline
\multicolumn{4}{r}{$\lambda^B_{ec}(\rm{LJ})$} \\
\cline{4-7}
Nuclide &$\lambda^0_{ec}$(FFN)  &$\lambda^0_{ec}$(AUFD) &$B_{12}=0$ &$B_{12}=10$ &$B_{12}=10^3$ &$B_{12}=10^6$ &$k1$ &$k2$\\
\hline
$^{55}$Mn   & 9.10e-3   &8.73e-3    &7.8895e-3  &1.3770e2   &3.1515e4   &2.7488e6  &1.1534 &1.1065  \\
$^{56}$Mn   & 9.29e-3   &2.90e-2    &2.6762e-2  &1.0681e2   &6.4156e5   &3.1205e6  & 0.8632&1.0836  \\
$^{57}$Mn   & 2.36e-5   &4.03e-4    &3.8320e-4  &3.7901e1   &2.1646e3   &1.9839e5  &0.1648 &1.0517\\
$^{58}$Mn   & 3.16e-4   &2.94e-3    &2.8113e-4  &5.8219e2   &1.3999e4   &1.3999e6  &0.1124 &1.0458\\
$^{59}$Fe   & 1.49e-4   &1.83e-4    &2.4663e-5  &8.9315e2   &7.7209e3   &1.5121e5  &0.6041 &0.7420\\
$^{61}$Ni   & ......    &5.07e-1    &3.3748e-1  &2.2129e2   &5.1440e4   &2.2142e6  &...... &1.5023 \\
$^{63}$Ni   & ......    &3.87e-3    &2.5660e-3  &1.2483e2   &2.2122e3   &1.3956e5  &...... &1.5082\\
\hline
\end{tabular}
\end{minipage}
\end{center}
\end{table*}
\begin{table*}
\caption{Comparisons of our calculations $\lambda^B_{ec}(\rm{LJ})$ in and not in SMFs for some typical iron group nuclides
with those of FFN's ($\lambda^0_{ec}(\rm{FFN}))$ and AUFD's ($\lambda^0_{ec}(\rm{AUFD}))$, which are for the case without SMFs at
$\rho_7=4010, Y_e=0.41, T_9=7.33$.}
\begin{center}
\tiny
\begin{minipage}{140mm}
\begin{tabular}{lllllllll}
\hline
\multicolumn{4}{r}{$\lambda^B_{ec}(\rm{LJ})$} \\
\cline{4-7}
Nuclide &$\lambda^0_{ec}$(FFN)  &$\lambda^0_{ec}$(AUFD) &$B_{12}=0$ &$B_{12}=10$ &$B_{12}=10^3$ &$B_{12}=10^6$ &$k1$ &$k2$\\
\hline
$^{57}$Mn   & 4.29e2    &8.36e2 &7.6335e2   &6.3917e4   &2.9624e8   &7.0466e10  &0.5620 &1.0952 \\
$^{58}$Mn   & 7.90e2    &1.05e3 &1.0326e3   &4.6822e5   &7.1383e9   &4.1146e11  & 0.7651&1.0169 \\
$^{59}$Mn   & ......    &1.41e2 &1.0132e2   &1.6386e3   &1.3991e8   &3.0604e10  &...... & 1.3916\\
$^{60}$Mn   & ......    &2.55e2 &1.1876e2   &1.8443e4   &2.6069e8   &2.9223e10  &...... &2.1472\\
$^{59}$Fe   & 7.43e2    &7.20e2 &6.1023e2   &4.5585e4   &5.7106e8   &5.7106e10  &1.2176 & 1.1799\\
$^{60}$Fe   & 1.441e1   &6.37e1 &2.6867e1   &1.0880e4   &1.0880e8   &3.3984e9   &0.5363 &2.3709\\
$^{61}$Fe   & ......    &1.63e2 &1.4785e2   &2.3397e3   &4.5281e9   &2.1010e10  &...... &1.1023\\
$^{56}$Cr   & ......    &3.33e1 &2.8845e1   &1.5366e3   &6.4200e9   &5.0816e9   &...... &1.1543\\
$^{57}$Cr   & ......    &6.09e1 &3.9436e1   &1.2695e3   &7.5182e9   &6.8590e9   &...... &1.5442\\
\hline
\end{tabular}
\end{minipage}
\end{center}
\end{table*}

In summary, one can see that the lower the temperature and density,
the less the effect on the EC rates from our numerical
calculations. This is because the GT transitions processes are
mainly aimed at in the high-energy region. Under the condition of
lower temperature, the average kinetic energy of electrons is
relatively small. The Fermi energy of electrons is also relatively
lower at low density. And most electrons will be in the weakly
degenerate state. The EC processes in the low--energy region make a
comparatively large contribution, while the high--energy GT
transition processes may not be dominated. On the contrary, the higher
the temperature, the larger the average energy of electrons, the
density and the Fermi energy of the electron gas become. The proportion
occupied by the GT transition increases and the contribution to the
rates may dominate.

According to above discussion, we can draw a conclusion that the SMFs
has a significant influence on the EC rates for a given
temperature-density point. Generally the higher the SMFs, the larger
affect on the EC rates become. The reason is that the Fermi energy
of electrons increases with the SMFs increasing and it may lead to more and more
electrons whose energy is above the threshold for EC process. The results
also show that the influences for different nuclei on EC reaction are
different due to the different Q--values and transition orbits in SMFs.

\begin{figure}
\centering
    \includegraphics[width=4cm,height=4cm]{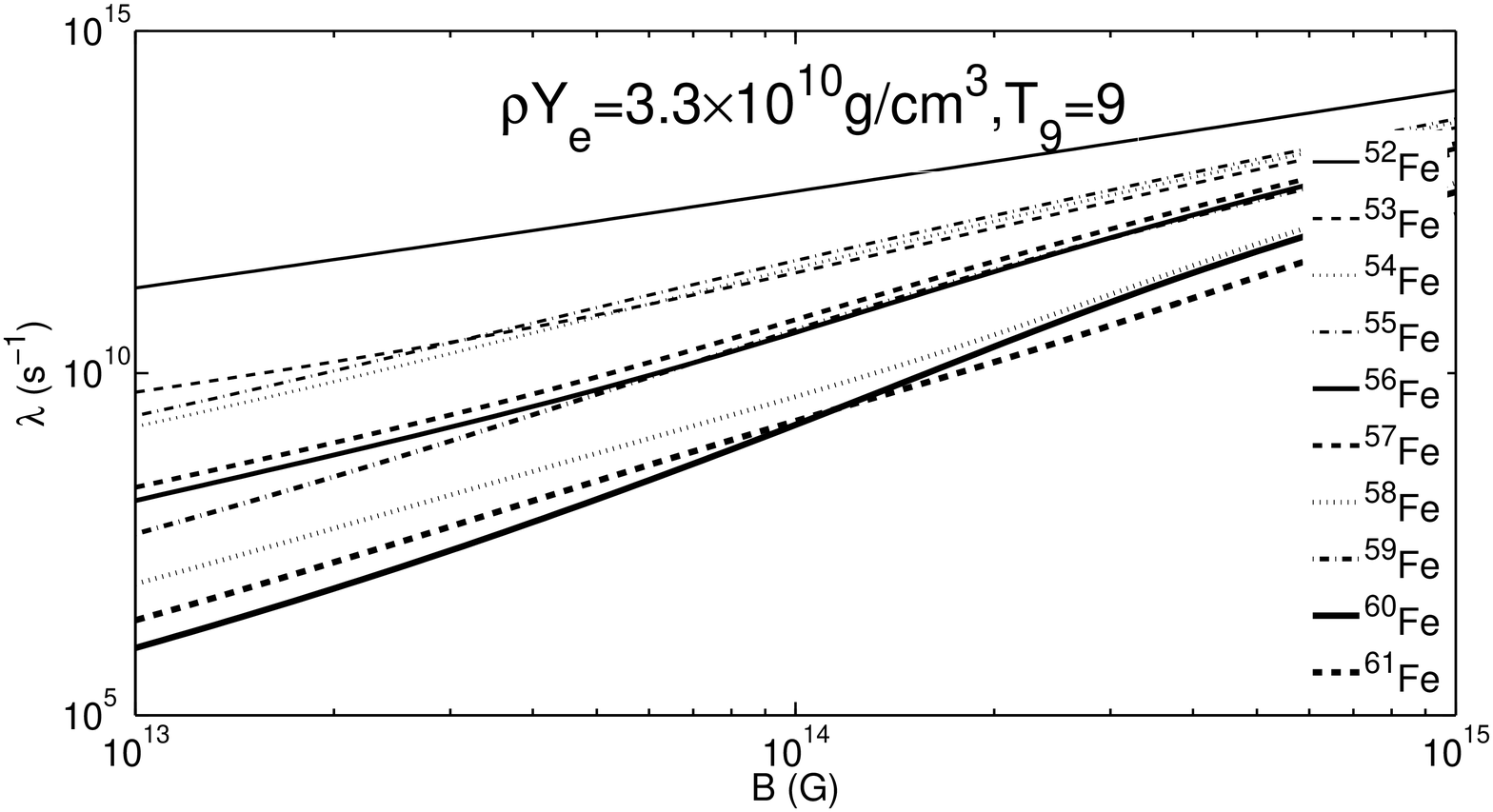}
    \includegraphics[width=4cm,height=4cm]{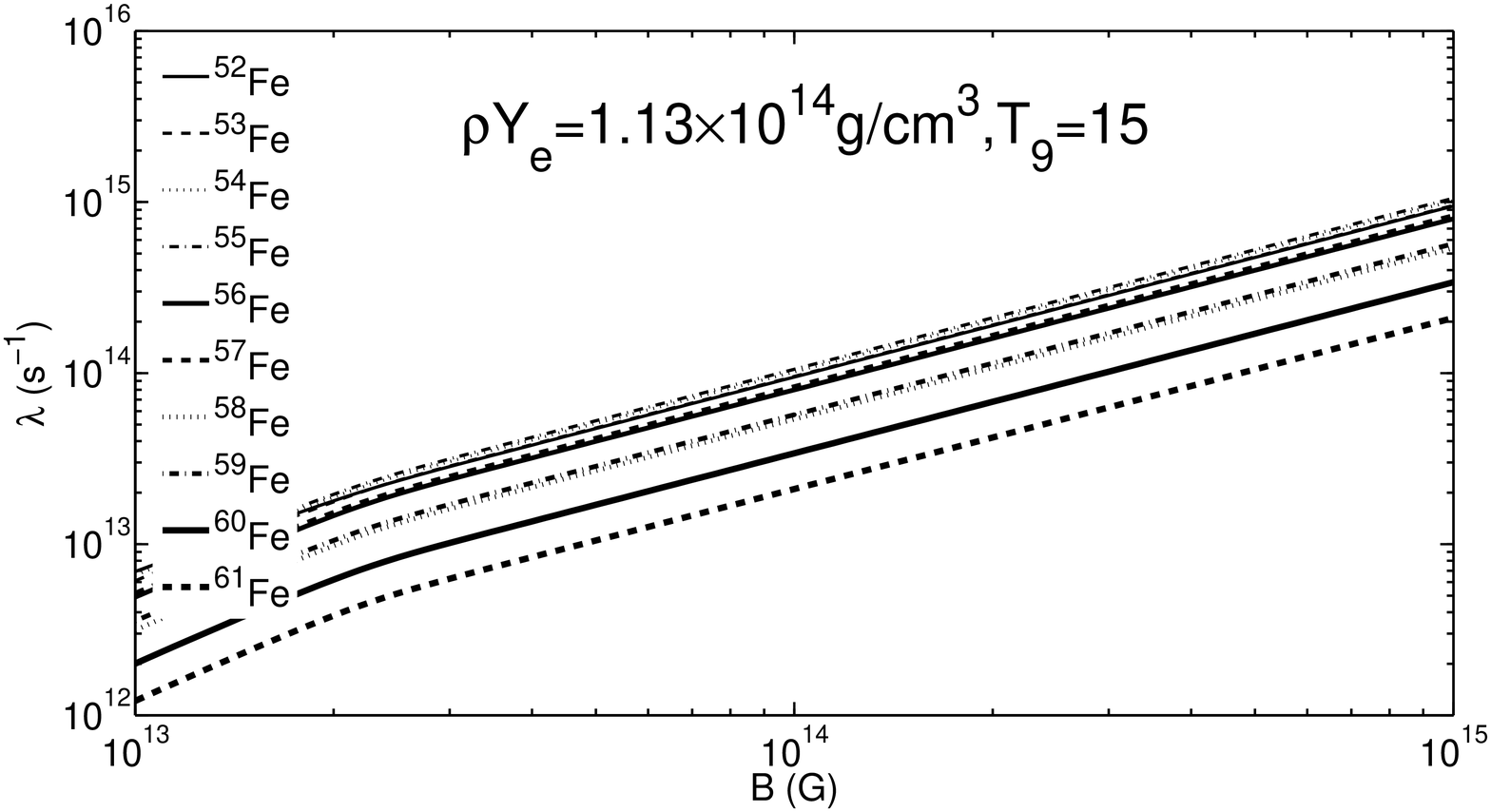}
   \caption{The EC rates for $^{52-61}$ Fe as a function of the magnetic field at the density
   and temperature of $\rho Y_e=3.3\times10^{10} g/cm^3$; $\rho Y_e=1.13\times10^{14} g/cm^3$ and $T_9=9; 15$ respectively}
   \label{Fig:1}
\end{figure}

%
\begin{figure}
\centering
    \includegraphics[width=4cm,height=4cm]{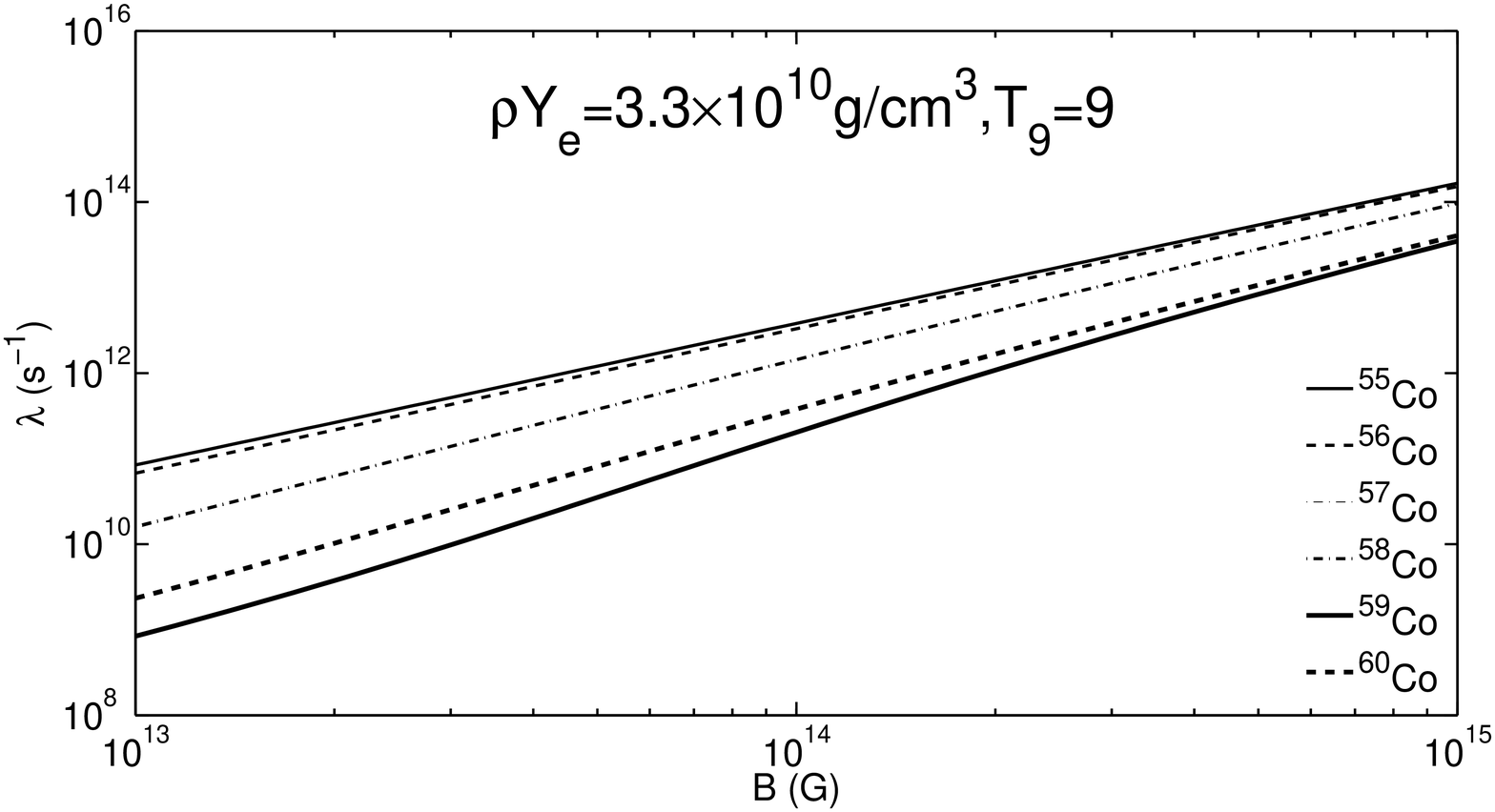}
    \includegraphics[width=4cm,height=4cm]{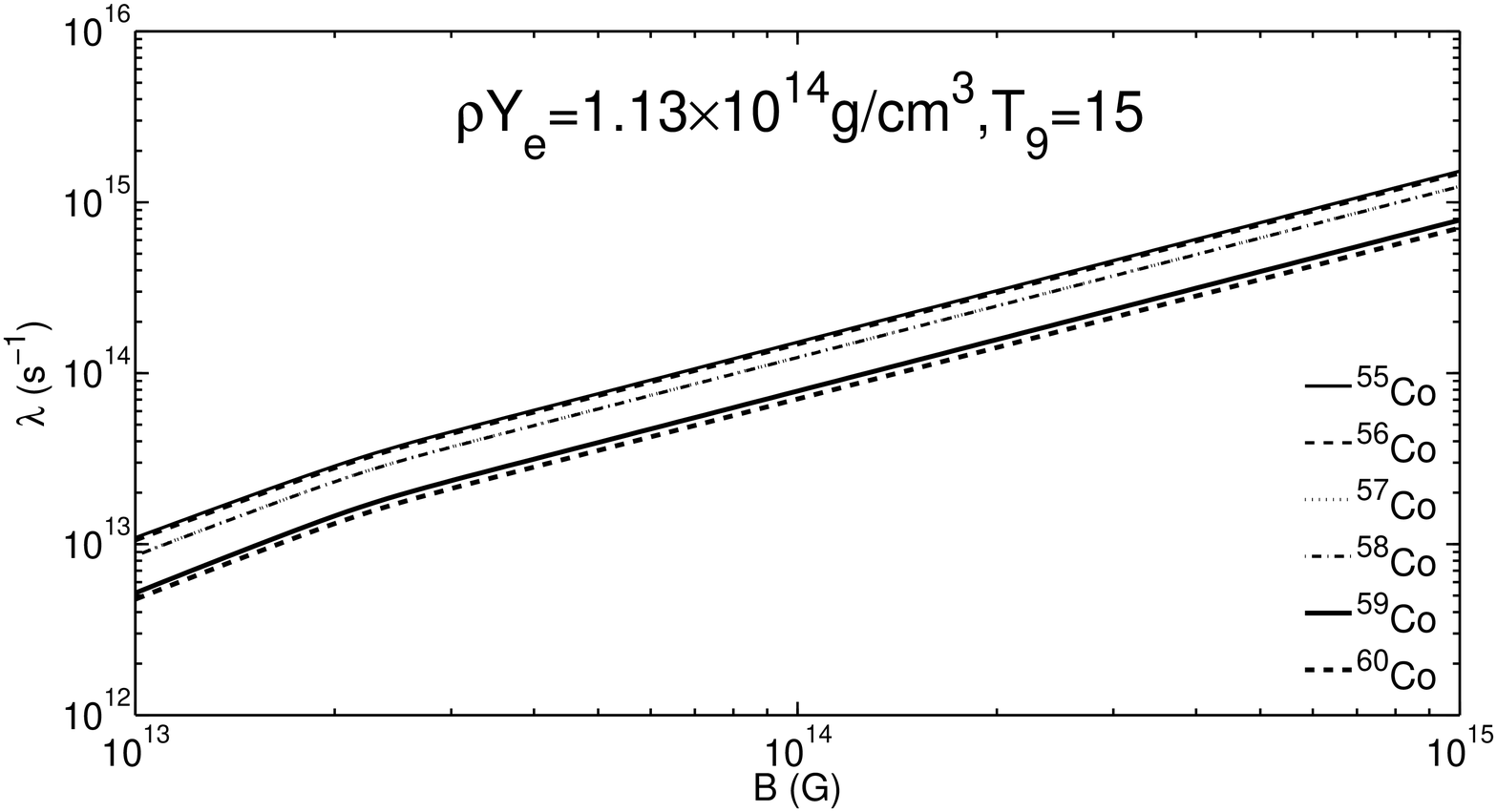}
   \caption{The EC rates for $^{55-60}$ Co as a function of the magnetic field at the density
   and temperature of $\rho Y_e=3.3\times10^{10} g/cm^3$; $\rho Y_e=1.13\times10^{14} g/cm^3$ and $T_9=9; 15$, respectively}
   \label{Fig:7}
\end{figure}
%
\begin{figure}
\centering
    \includegraphics[width=4cm,height=4cm]{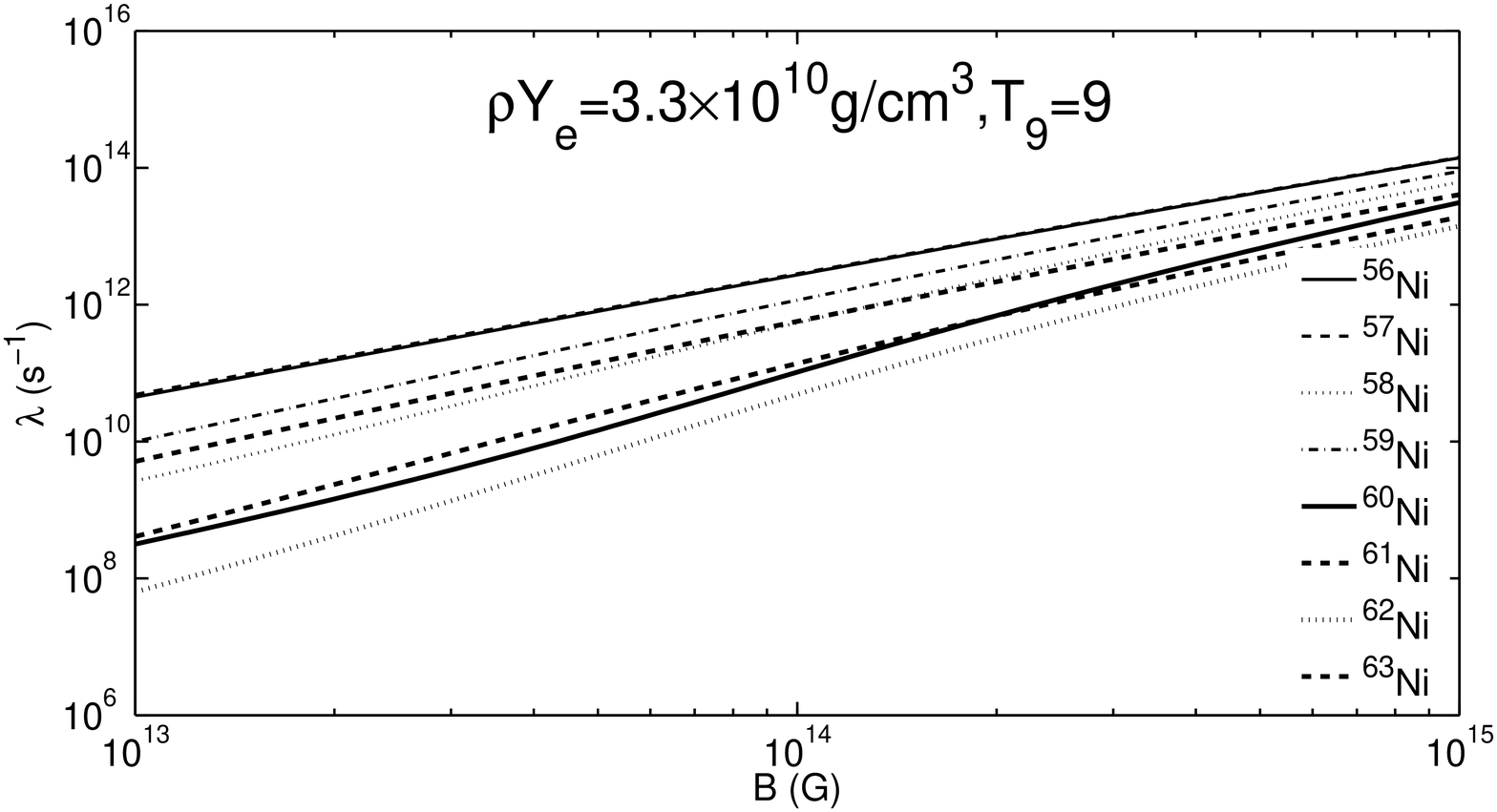}
    \includegraphics[width=4cm,height=4cm]{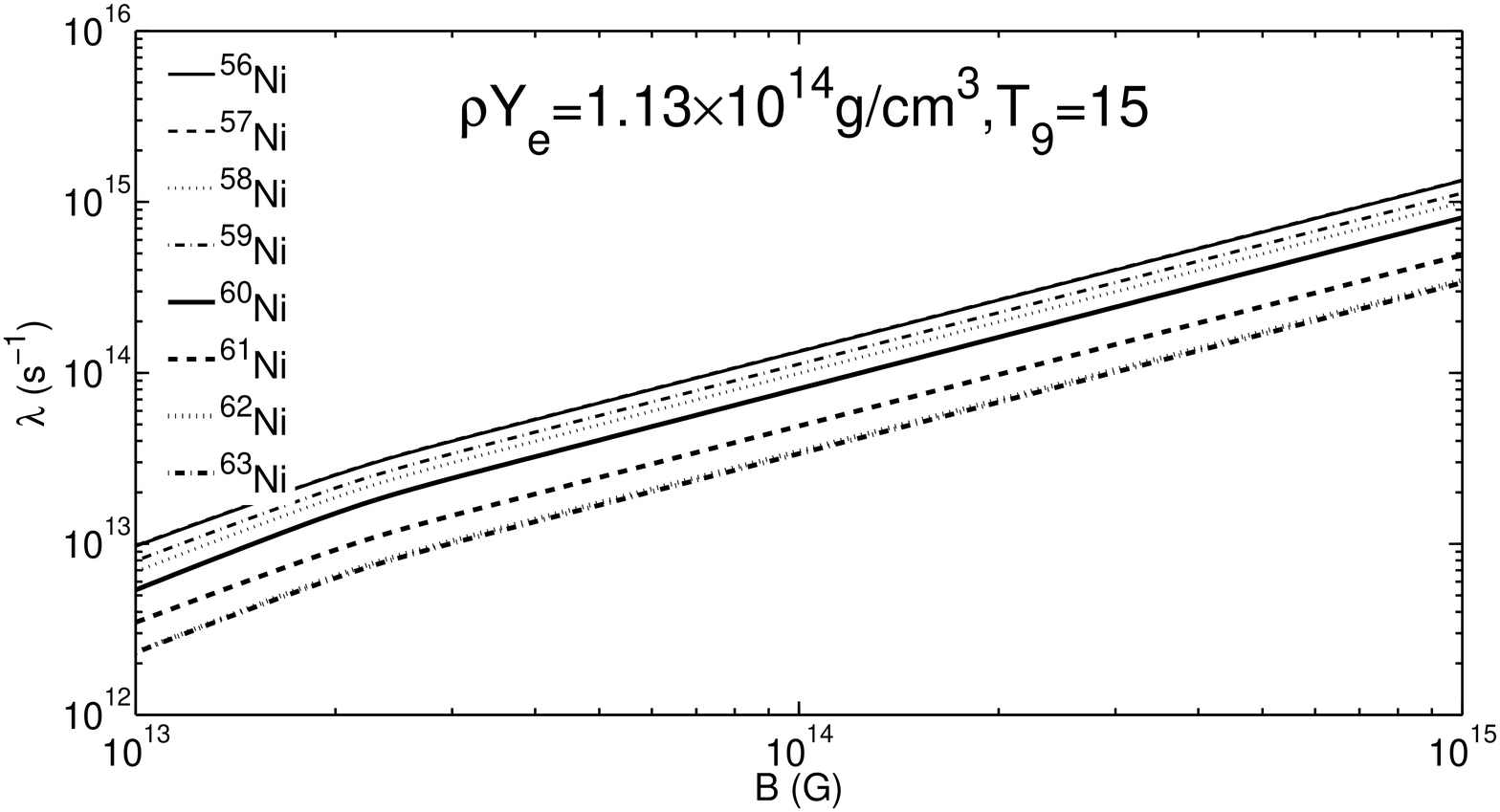}
   \caption{The EC rates for $^{56-63}$ Ni as a function of the magnetic field at the density
   and temperature of $\rho Y_e=3.3\times10^{10} g/cm^3$; $\rho Y_e=1.13\times10^{14} g/cm^3$ and $T_9=9; 15$ respectively }
\label{Fig:8}
\end{figure}

\begin{figure}
\centering
    \includegraphics[width=4cm,height=4cm]{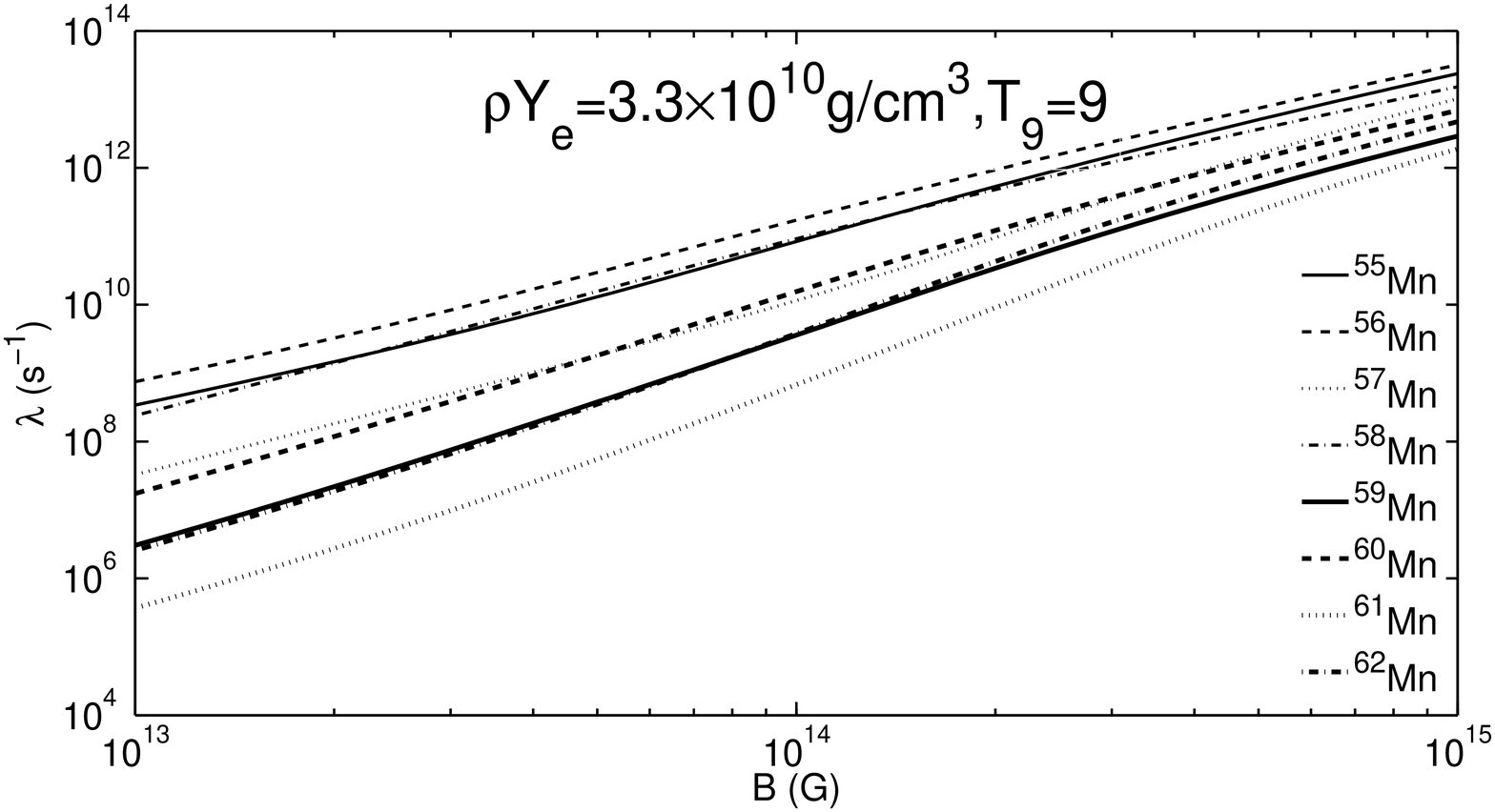}
    \includegraphics[width=4cm,height=4cm]{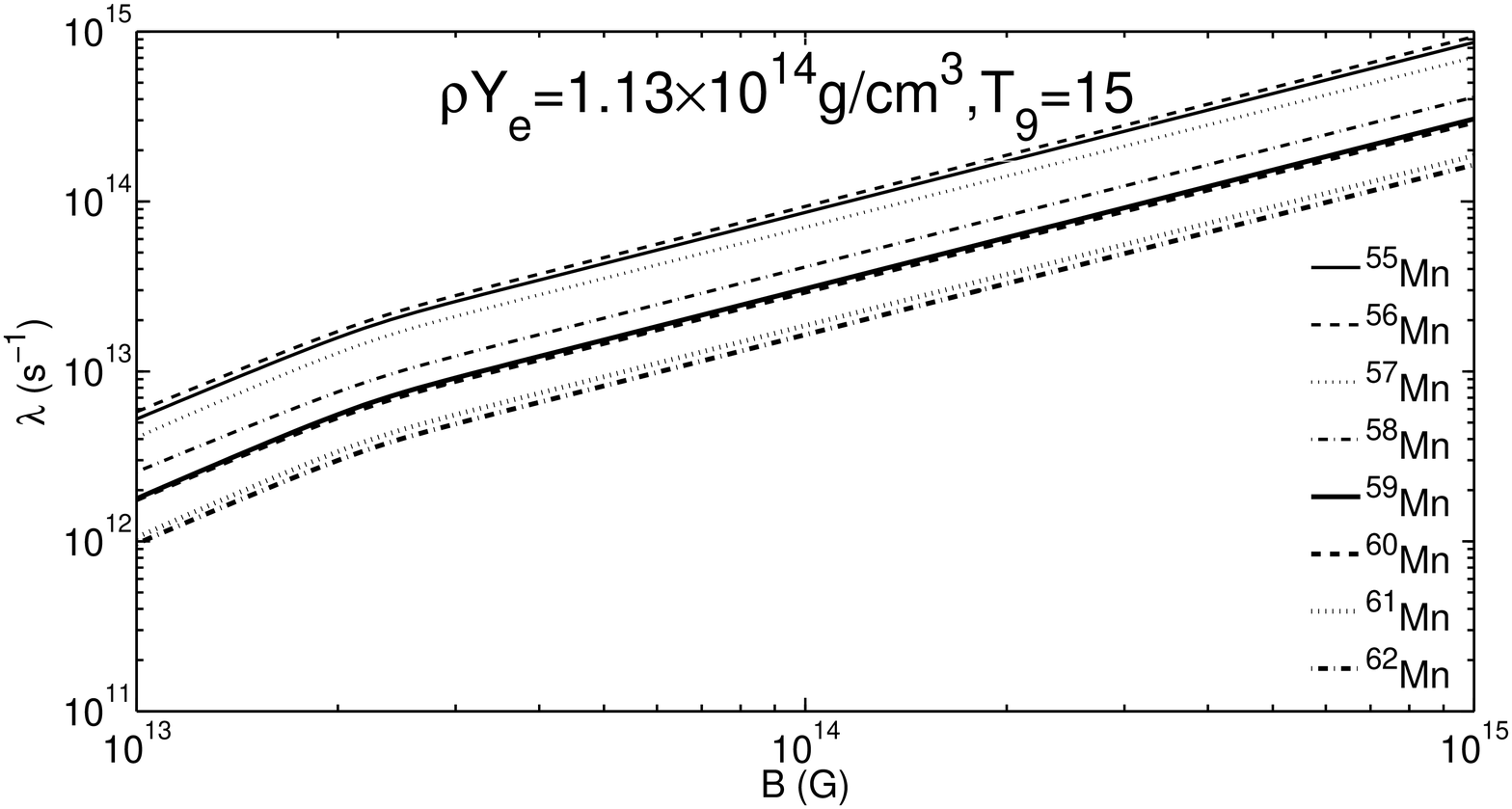}
   \caption{  The EC rates for $^{55-62}$ Mn as a function of the magnetic field at the density
   and temperature of $\rho Y_e=3.3\times10^{10} g/cm^3$; $\rho Y_e=1.13\times10^{14} g/cm^3$ and $T_9=9; 15$ respectively}
   \label{Fig:9}
\end{figure}

\begin{figure}
\centering
    \includegraphics[width=4cm,height=4cm]{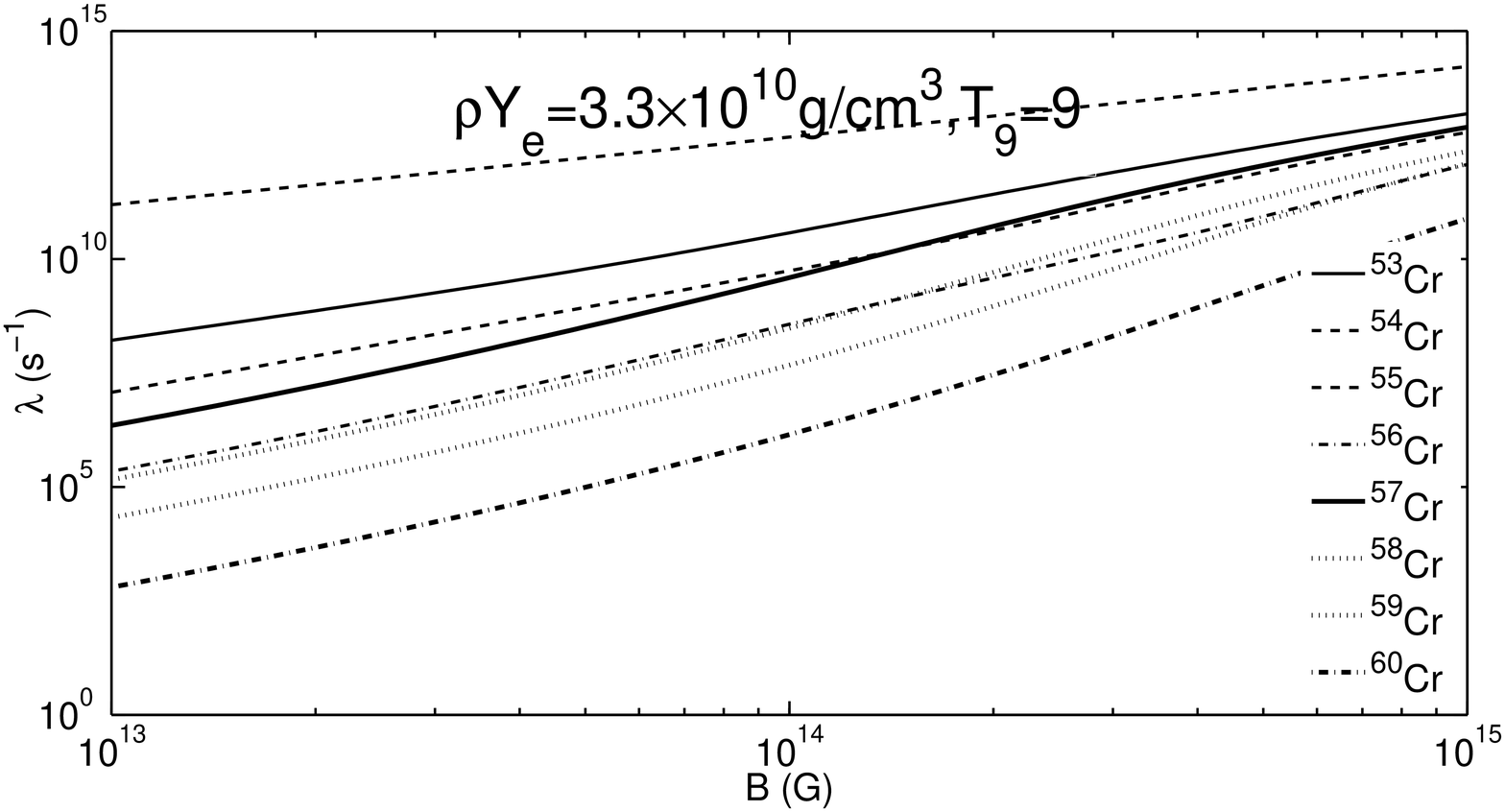}
    \includegraphics[width=4cm,height=4cm]{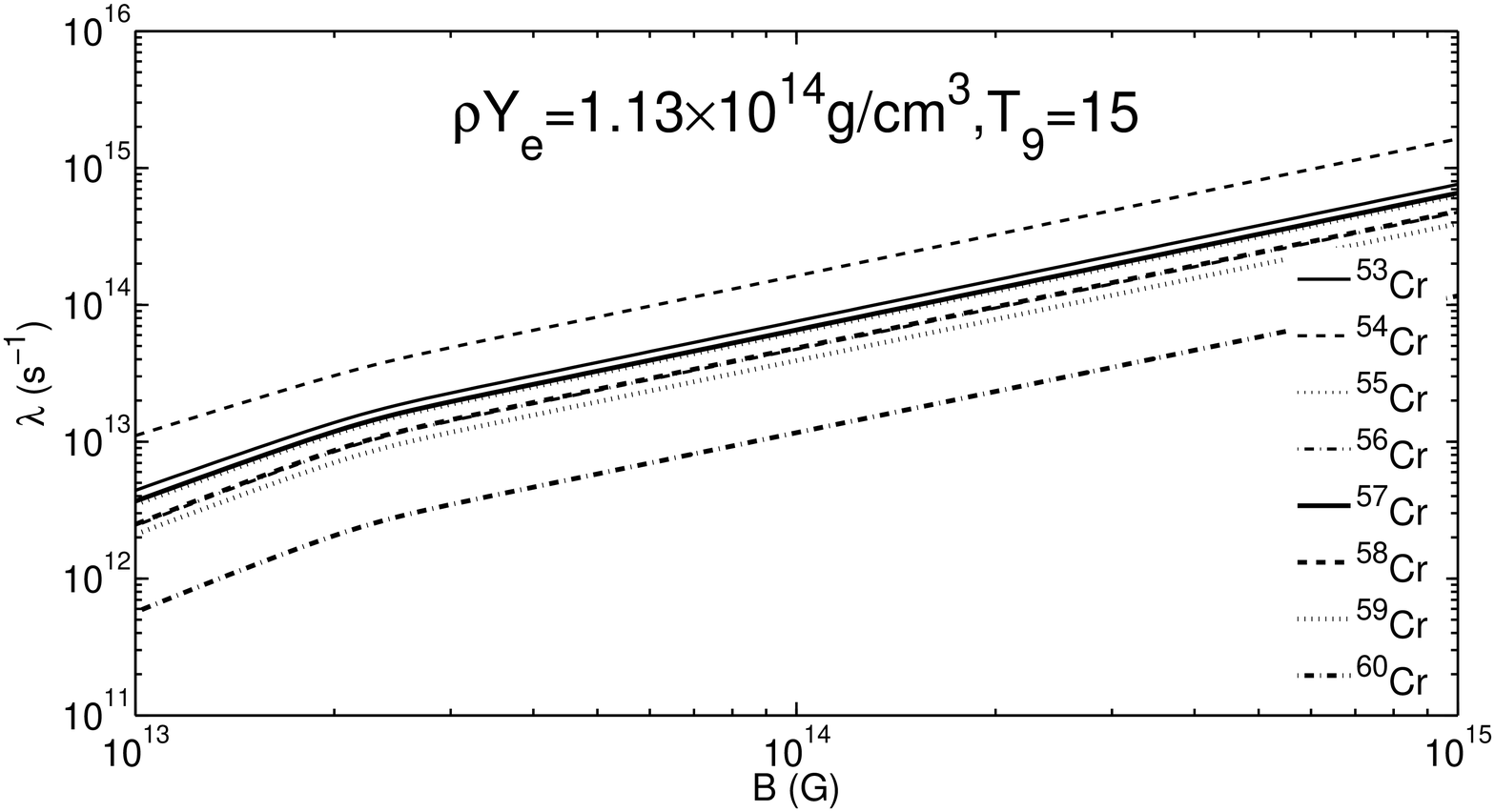}
   \caption{  The EC rates for $^{53-60}$ Cr  as a function of the magnetic field at the density
   and temperature of $\rho Y_e=3.3\times10^{10} g/cm^3$; $\rho Y_e=1.13\times10^{14} g/cm^3$ and $T_9=9; 15$ respectively}
   \label{Fig:10}
\end{figure}
\begin{figure}
\centering
    \includegraphics[width=4cm,height=4cm]{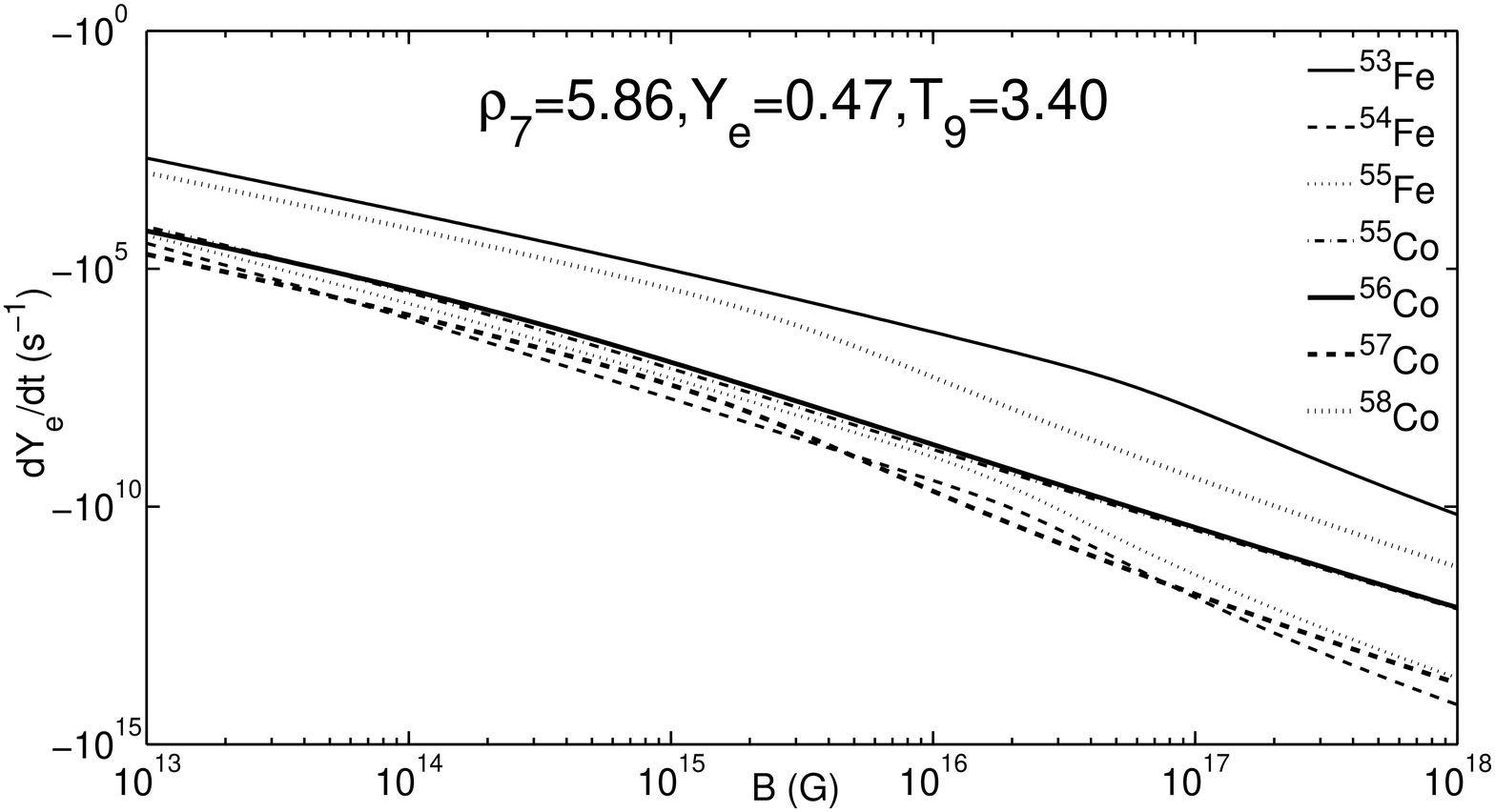}
    \includegraphics[width=4cm,height=4cm]{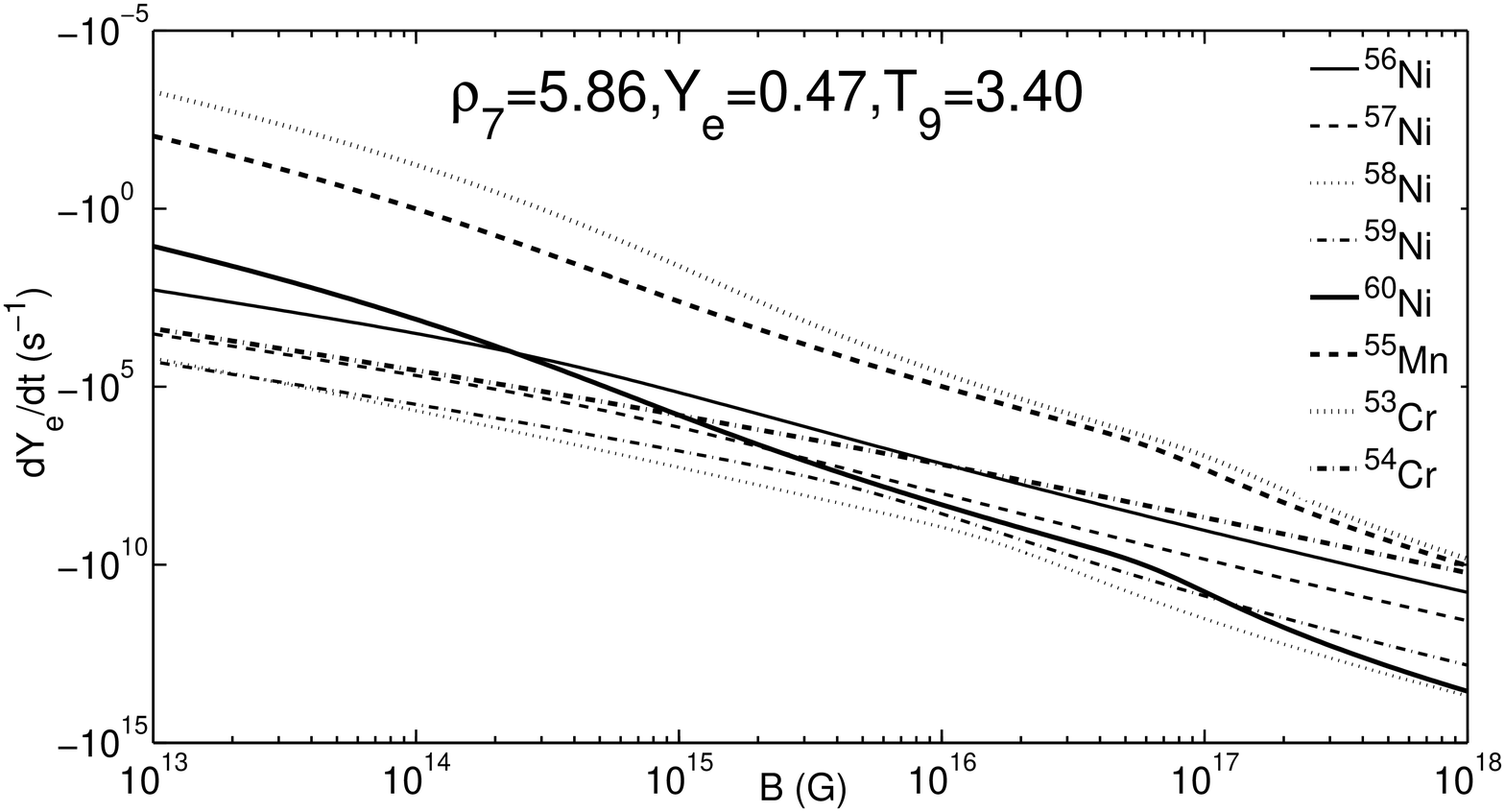}
   \caption{The RCEA  as a function of the magnetic field at the density  and temperature of $\rho_7=5.86, Y_e=0.47$; and $T_9=3.40$}
\end{figure}

\begin{figure}
\centering
    \includegraphics[width=4cm,height=4cm]{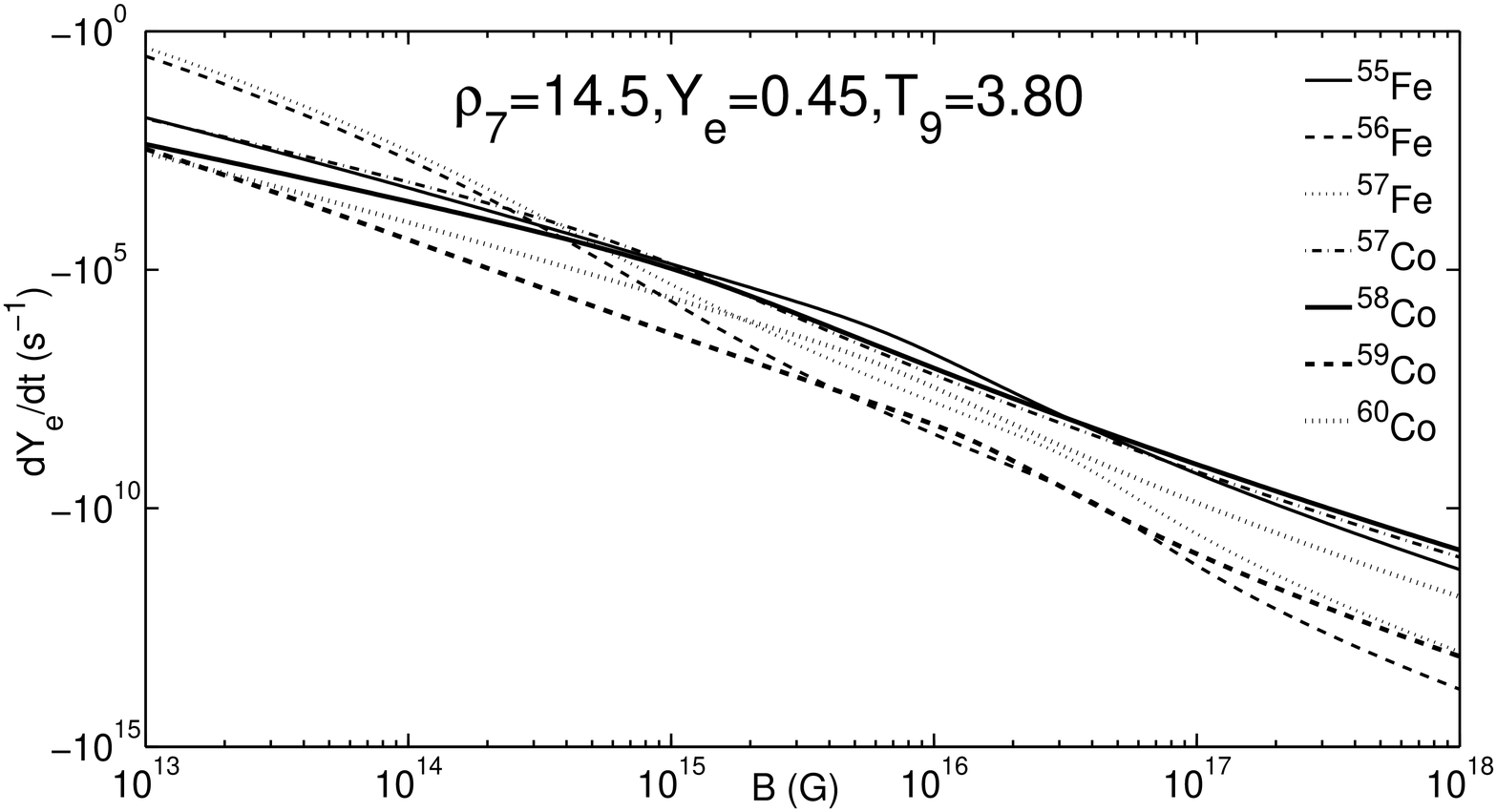}
    \includegraphics[width=4cm,height=4cm]{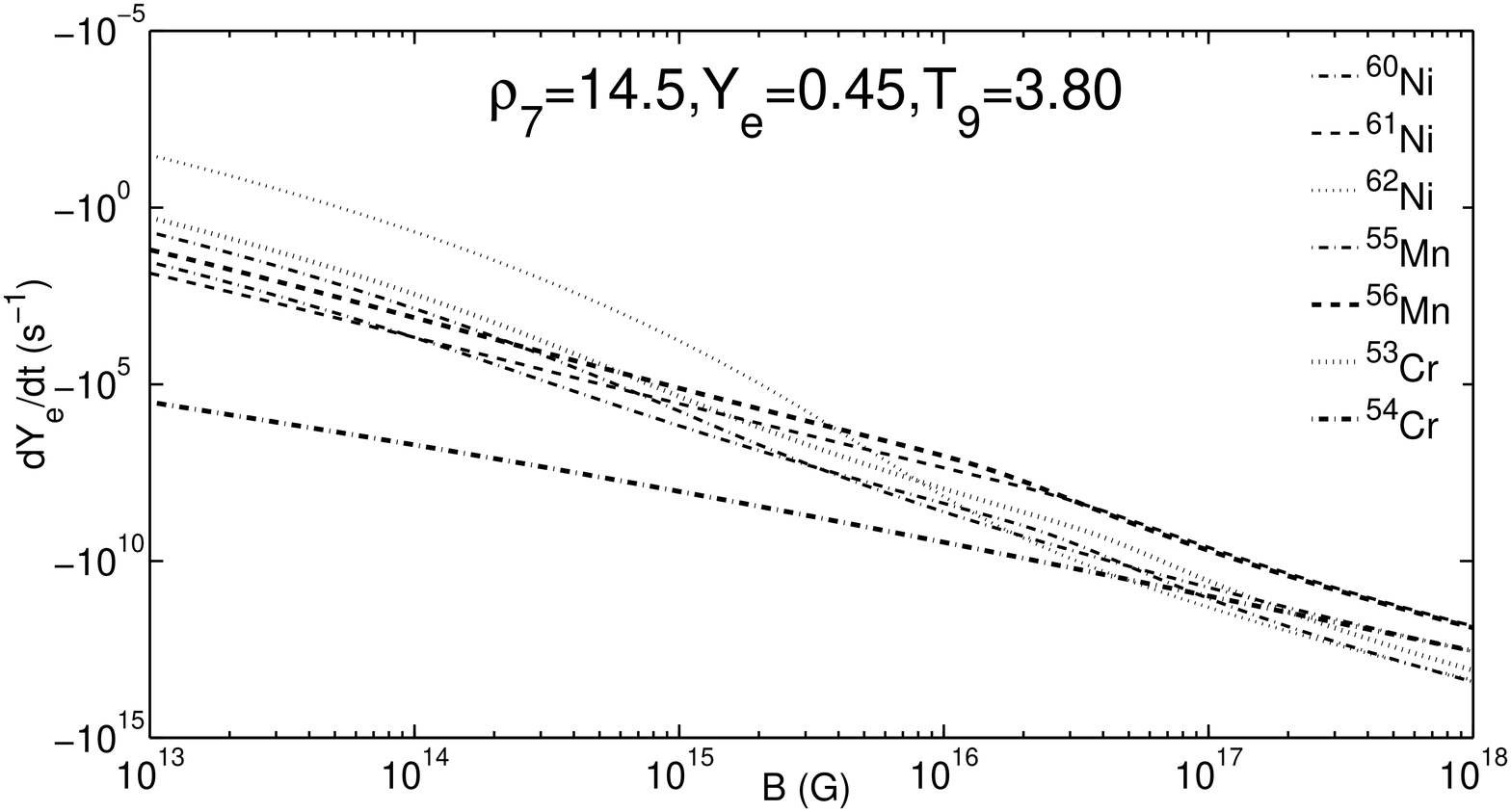}
   \caption{The RCEA  as a function of the magnetic field at the density  and temperature of $\rho_7=14.5, Y_e=0.45$; and $T_9=3.80$}
   \label{Fig:4}
\end{figure}

\begin{figure}
\centering
    \includegraphics[width=4cm,height=4cm]{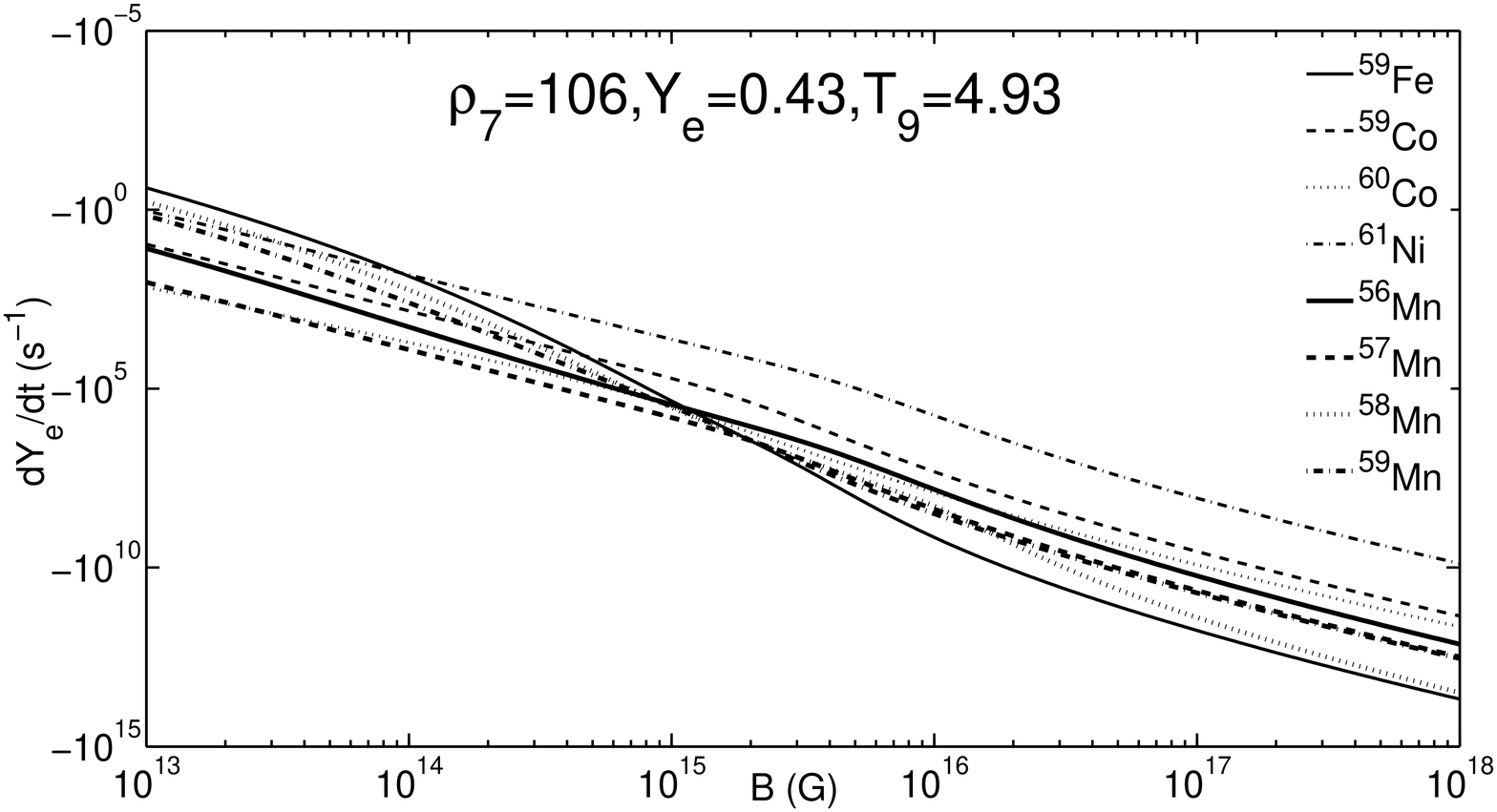}
    \includegraphics[width=4cm,height=4cm]{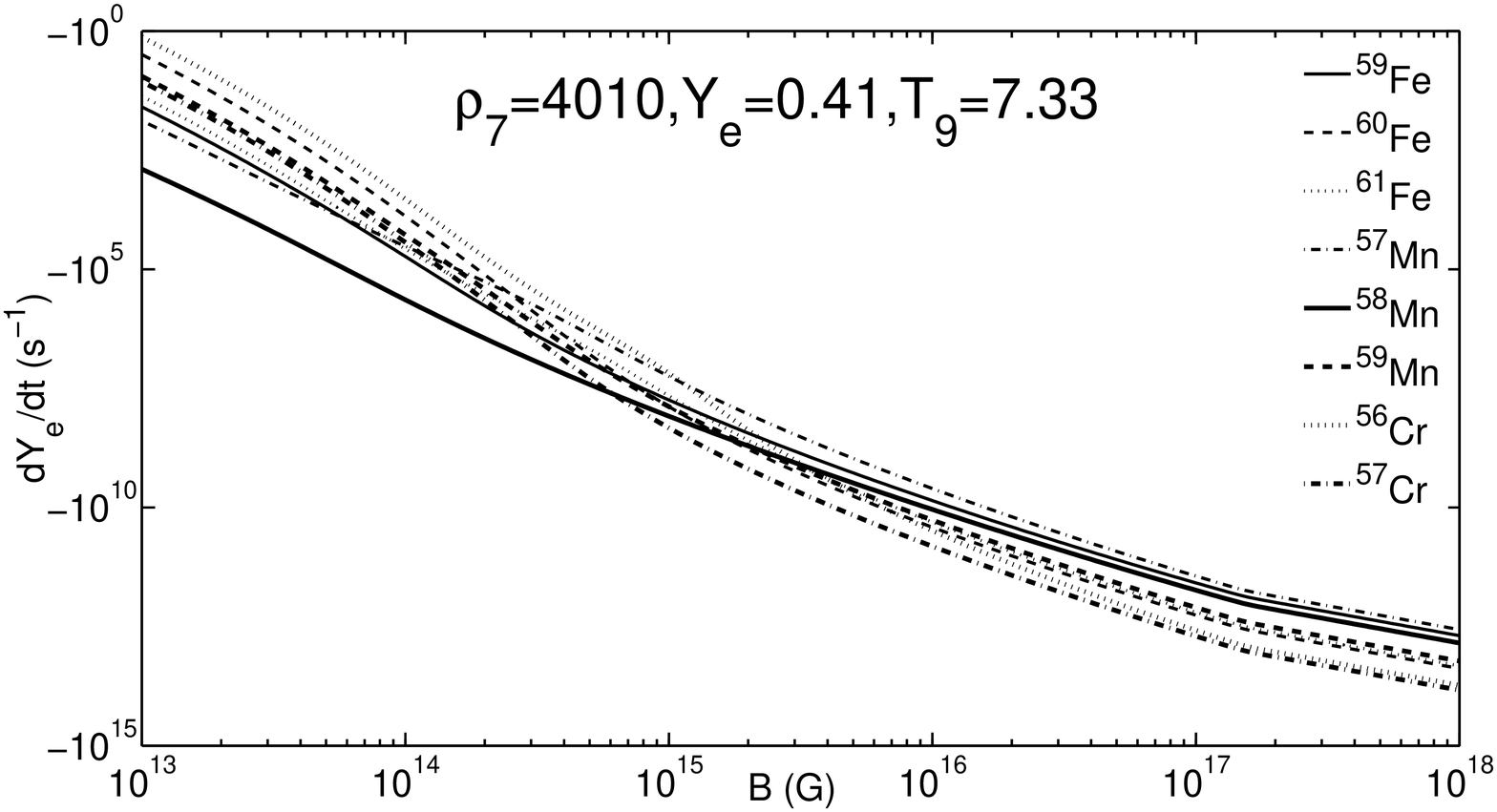}
   \caption{The RCEA  as a function of the magnetic field at the density  and temperature of
   $\rho_7=106, Y_e=0.43$; $T_9=7.93$ and $\rho_7=4010, Y_e=0.41$;  $T_9=7.33$ respectively}
   \label{Fig:5}
\end{figure}

As everyone knows, what really matters for stellar evolution is electronic abundance $Y_e$.
Electronic abundance variation is one of crucial parameters in supernova.
The electronic abundance serious influences the changes of electron degenerate pressure and entropy at
late stages of stellar evolution, especially at the supernova explosion process and plays a very important role.
Figures 11-13 show the RCEA due to EC of iron group nuclei as a function of magnetic field in SMFs.
As seen in the three figures, SMFs have a large impact on the RCEA for most iron group nuclei. The $Y_e$ due to SMFs may be almost reduced
more than eight orders of magnitude.

From above Figures, we can see the change of the electron fraction is very sensitivity parameter in EC process.
The main reason resulted in the EC rates are increased greatly by SMFs. It may be induce the number of the electrons reduce
largely. With increasing of the density and the magnetic field, the electron chemical potential becomes so high that
rapid and great progress would be made due to large numbers of electrons join in EC reaction.

From the oxygen shell burning phase up to the end of convective core silicon burning phase of massive stars the EC rates on these
nuclides play important roles. FFN had done some pioneer works on EC rates. In order to understand how much EC is affected by SMFs,
the comparisons of our results $\lambda^B_{ec}$(LJ)in SMFs with those of FFN's \citep{b8, b9}($\lambda^0_{ec}$(FFN)) and
AUFD's \citep{b1}($\lambda^0_{ec}$ (AUFD)) for the case without SMFs are presented in tabular forms. Table 1-4 show the comparison of
our results in SMFs with those FFN's and AUFD's in different astrophysical environments.

The calculated rates for most nuclides due to SMF are increased and even exceeded as much as for nine orders of magnitude of compared to FFN's result,
which for the case without SMFs. The four tables also show the comparisons of our results in SMFs with those and AUFD's, which are for the case
without SMFs. The calculated rates for most nuclides due to SMFs are increased and even exceeded as much as for eight orders of magnitude of compared
to AUFD's result.

For the case without SMFs the four tables show the comparison of
calculated EC rates of our results with those FFN's and AUFD's. We
find that our results are in well agreement with AUFD's from the
scale factor $k_1$ and $k_2$, but FFN's are about close to one order
magnitude bigger than ours.

The question about the blocking and thermal unblocking of final
states in the process of EC, has discussed in detail by \citet{b12}.
\citet{b10} also show that all the corresponding neutron orbits are
filled and the valence neutrons are in the $gd$--shell during
collapse a point is reached where a average nucleus is so
neutron-rich that, while the valence protons are still filling
$pf$--shell orbits \citep{b29}. However, we discuss the Gamow
-Teller strength in stellar electron capture process on iron group
nuclides based on $pf$--shell model in this paper. As an example,
for $^{58}$Co in process of EC, a proton is turned into a neutron by
a GT transition and Fermi transition is blocked. All seven $pf$
protons are filled in the $1f_{7/2}$ orbital and thus the only
allowed GT transitions are to the $1f_{7/2}$ and $1f_{5/2}$ neutron
orbitals. The formal orbital is blocked by neutrons. Thus a
$1f_{7/2}$ proton has become a $1f_{5/2}$ neutron.

On the other hand, the charge exchange reactions (p, n) and (n, p)
make it possible to observe, in principle, the total GT strength
distribution in nuclei. For some iron nuclides the experimental
information is particularly rich  and the availability of both
$\rm{GT}^+$ and $\rm{GT}^-$. It makes possible to study in detail
the problem of renormalization of $\sigma\tau$ operators. The total
calculated GT strength in a full $pf$--shell calculation, resulting
in $\rm{B}(\rm{GT})=g_A^2|\langle\vec{\sigma}\tau_{+}\rangle|^2$,
where $g_{\rm{A}}^2$ is axial-vector coupling constant\citep{b13}.

For example, the electron capture on $^{56}$Ni is dominated by the
wave functions of the parent and daughter states and effected
greatly by SMFs. And the total GT strength for $^{56}$Ni in a full
$pf$--shell calculation, resulting in $\rm{B}(\rm{GT})=10.1 g_A^2$.
The total GT strength of the other important nuclide $^{56}$Fe in a
full $pf$--shell calculation can be found in the Ref. \citep{b5}. An
average of the GT strength distribution is in fact obtained by SMMC
method.

The electron capture of the neutron rich nuclide do not has measured
mass, so that the EC Q-value has to be estimated with a mass formal
by FFN. FFN used the \citet{b30} Semiempirical atomic mass formula,
so that the Q-value used in the effective rates are quite different.
On the basis of the independent particle model, FFN parametrized the
GT contribution to the EC rates. To complete the FFN rate estimate,
the GT contribution has been supplemented by a contribution
simulating low-lying transitions. Basing on nuclear shell model,
AUFD expanded the FFN's works and analyzed the nuclear excited level
by a Simple calculation on the nuclear excitation level transitions
in their works. The capture rates are made up of the lower energy
transition rates between the ground states and the higher energy
transition rates between GT resonance states. Their work simplifies
the nuclear excited energy level transition calculation, Thus the
calculation method is a little rough.

It is generally known that the EC rate is easily calculated as long as the distribution of nuclear
 excited states is clear. But it is very difficult to know the distribution because a lot of nuclear
 excited states are not stable at the stage of supernova explosive. The experimental data on GT distribution
 within the nucleus becomes available. These data show that the EC value of the FFN's works \citep{b8, b9},
 which using the GT contribution parameters have large errors. These data clearly show that the GT strength disappear
 in the daughter nucleus and split into several excitation. AUFD analyze the nuclear excited level by a Simple calculation
 on the nuclear excitation level transitions in their works. The excited state is splinted up into the ground state near the
 low energy parts and resonant states near the high-energy parts respectively. Therefore, the capture rates are made up of the
 lower energy transition rates between the ground states and the higher energy transition rates between GT resonance states.
 Therefore the works of AUFD is an oversimplification and the accuracy is limited.

\section{Conclusions}
As everyone knows, the problems  about the influence of a strong magnetic field on electron capture in non-zero temperature
crusts of neutron stars have already been discussed by \citet{b7}; \citet{b14}. Their results show that the magnetic fields ($10^{9} - 10^{13}G$) on surfaces of neutron stars have almost no effect on the electron capture rate. In this paper, we have carried out estimation for EC rates and RCEA of iron group nuclei in SMFs which is ranging from $10^{13}G$ to $10^{18}G$. It is concluded that an SMFs has a significant effect on them For most iron group nuclides.

We also compare our results with those of FFN's and AUFD's for the
case with and without SMFs. The results show that our results is in
well agreement with AUFD's, but the FFN's are about close to one
order magnitude bigger than ours for the case without SMFs. On
contrary, our calculated rates for most nuclides in SMFs are
increased and even exceeded as much as by nine and eight orders of
magnitude bigger than FFN's and AUFD's, which are in case without
SMFs, respectively.

As is known to all, the research on EC reaction at late stages of stellar evolution will still be a long-term and
arduous task. It will have a key and profound significance to better understand and study the final evolutionary paths
and nuclear interaction model of cosmic objects, as well as the physical structure and state in the interiors of celestial
objects in the environment of high temperature, high pressure, high magnetic fields and high density. Our conclusions may be
helpful to the investigation of the late evolution of the neutron stars and magnetars, the nucleosyntheses of heavy elements
and the numerical calculations of stellar evolution.

\section*{Acknowledgments}

This work was supported by the Advanced Academy Special Foundation of Sanya under Grant No 2011YD14.

\bsp

\label{lastpage}

\end{document}